\documentclass[aps,prl,superscriptaddress,amsmath,amssymb]{revtex4-2}

\usepackage{comment}
\usepackage{tikz}
\usetikzlibrary{trees}
\usetikzlibrary{decorations.pathmorphing}
\usetikzlibrary{decorations.markings}
\tikzset{
    photon/.style={decorate, decoration={snake,segment length=1.5mm}, draw=black},
    coulomb/.style={dotted},
    electron/.style={draw=black, postaction={decorate},
        decoration={markings,mark=at position .55 with {\arrow[draw=black]{>}}}}, 
    gluon/.style={decorate, draw=magenta,
        decoration={coil,amplitude=4pt, segment length=5pt}},
    boundelectron/.style={thick, double},
    transverse/.style={dashed}
}

\usepackage{graphicx}
\usepackage{dcolumn}
\usepackage{bm}
\usepackage{rotating}
\usepackage{multirow}

\newcolumntype{.}{D{.}{.}{14}}

\usepackage[utf8]{inputenc}
\usepackage{physics}
\usepackage{amsmath, amssymb}
\usepackage{tabularx,booktabs,array,dcolumn}
\usepackage{setspace}
\usepackage{vmargin}
\usepackage{xcolor}
\usepackage{graphicx,psfrag,subfigure}

\usepackage{float}
\usepackage[all]{xy}

\usepackage{bm}
\usepackage{mathtools}
\usepackage{physics}
\usepackage[english]{babel}

\newcommand{\bos}[1]{\boldsymbol{#1}}
\newcommand{\mr}[1]{\mathrm{#1}}
\newcommand{\pd}[2]{\frac{\partial #1}{\partial #2}}

\def\Eh{E_\mathrm{h}}

\def\iim{\mr{i}}
\def\eem{\mr{e}}

\def\DC{\text{DC}}
\def\DCB{\text{DCB}}
\def\DCpB{{\text{DC}\langle \text{B}\rangle}}

\def\nb{N_\text{b}} 

\def\unittwo{1^{[2]}} 
\def\unitfour{1^{[4]}} 

\def\four{^{[4]}} 

\def\dd{\mathrm{d}}

\def\bR{\bos{R}}

\def\bp{\bos{p}}
\def\br{\bos{r}}

\def\bsigma{\bos{\sigma}}

\def\epsi{\varepsilon}

\def\mL{\Lambda}

\def\som{Supplementary Material}

\def\texte{\text{e}}

\def\pos{\lbrace \text{e}^-,\text{e}^+\rbrace}
\def\muo{\lbrace \text{e}^-,\mu^+\rbrace}
\def\hyd{\lbrace \text{e}^-,\text{p}^+\rbrace}
\def\muhyd{\lbrace \mu^-,\text{p}^+\rbrace}
\def\muH{$\mu$H}

\def\DC{\text{DC}}
\def\DCpB{\text{DC}\langle\text{B}\rangle}
\def\DCptwoB{\text{DC}\mathcal{B}_2}
\def\DCB{\text{DCB}}
\def\pupu{{\scalebox{0.65}{$++$}}}

\def\dd{\mathrm{d}}

\def\bp{\bos{p}}
\def\br{\bos{r}}

\def\bsigma{\bos{\sigma}}

\newcommand{\mxelmn}[3]{\matrixel{#1 f_\mu}{#2}{#3 f_\nu}}

\usepackage[unicode]{hyperref}
\usepackage{soul}

\newcolumntype{d}[1]{D{.}{.}{#1}}

\usepackage[unicode]{hyperref}
\hypersetup{
   unicode=true,          
   plainpages=false,
   colorlinks=true,       
   linkcolor=blue,          
   citecolor=blue,        
}

\usepackage{url}

\begin{document}

\title{%
Pre-Born--Oppenheimer Dirac--Coulomb--Breit computations for two-body systems
}

\author{D\'avid Ferenc}
\author{Edit M\'atyus} 
\email{edit.matyus@ttk.elte.hu}
\affiliation{ELTE, Eötvös Loránd University, Institute of Chemistry, 
Pázmány Péter sétány 1/A, Budapest, H-1117, Hungary}

\date{\today}

\begin{abstract}
\noindent %
The sixteen-component, no-pair Dirac--Coulomb--Breit equation, derived from the Bethe--Salpeter equation, is solved in a variational procedure using Gaussian-type basis functions for the example of positronium, muonium, hydrogen atom, and muonic hydrogen. 
The $\alpha$ fine-structure-constant dependence of the variational energies, through fitting a function of $\alpha^n$ and $\alpha^n\text{ln}\alpha$ terms, shows excellent agreement with the relevant energy expressions of the (perturbative) non-relativistic QED framework, and thereby, establishes a solid reference for the development of a computational relativistic QED approach.
\end{abstract}

\maketitle

\noindent%
The positronium, $\text{Ps}=\pos$, muonium, $\text{Mu}=\muo$, hydrogen atom, $\text{H}=\hyd$, and muonic-hydrogen, $\mu\text{H}=\muhyd$, are the simplest, yet some of the most extensively studied bound-state systems. 
Their simplicity allows for the high-precision evaluation of energy corrections arising from
special relativity and interactions from the matter and photon fields \cite{BeSabook57,EiGrSh01}.
The high-precision spectroscopy experiments \cite{FeMiChChDaChZu93,Ha06,Bi09,hatom13,IsNaAsKoSaYTaYa14,FrPePe19,GuBaHoCa20,MuMASS22} together with the theoretical results (see Refs.~\cite{EiGrSh01,AdCaPe22} and references therein) provide stringent test for validity of quantum electrodynamics (QED) in the low-energy range and probe physics beyond the Standard Model \cite{Ru04,Karshenboim04,Karshenboim05,GnKsMaRu06,rmp18}.
Ps is a candidate for precision free-fall experiments to test QED and gravity \cite{Ka16},
H and \muH\ are the stars of the famous proton-size puzzle \cite{protonsize17,FlGaThBoJuBiNeAbGu18,KaMa2019},
while Mu has attracted interest in relation with the muon's anomalous magnetic moment \cite{Cr18,MuMASS22}.

For bound-state systems, it is relevant to have a wave equation that can be solved to obtain a good zeroth-order description. 
So far, the non-relativistic Schrödinger equation has been used as reference, which has analytic solution for two-body systems. Then, relativistic and QED corrections have been derived corresponding to increasing orders of the $\alpha$ fine-structure constant.
We call these corrections, for short, non-relativistic QED (nrQED) corrections.
A recent review \cite{AdCaPe22} provides an excellent overview of the extensive literature of higher-order nrQED corrections to positronium energies. Corrections up to $\alpha^6$ order (in natural units, $\alpha^4\Eh$ in hartree atomic units) are considered complete, and ongoing work is about $\alpha^7$ order corrections. Some of the calculations have been carried out not only for equal but arbitrary spin-1/2 fermion masses.

In the present work, we do not aim to reproduce the formally derived nrQED expressions, but initiate an alternative approach to the two-particle relativistic QED problem based on a zeroth-order wave equation in which special relativity is already accounted for. 
The theoretical framework for this (computational) relativistic QED program is provided by the Bethe--Salpeter equation \cite{SaBe51}, derived from field theory \cite{GMLo51}, and its Salpeter--Sucher exact equal-time form \cite{Sa52,sucherPhD1958}, which provides us a no-pair, two-particle relativistic wave equation, 
\begin{align}
  (H + H_\Delta) \Psi
  =
  E \Psi \; ,
  \label{eq:eBS}
\end{align}
which has the form of a Schrödinger-like wave equation, for which high-precision numerical solution techniques can be adapted.
The $\Psi$ wave function in Eq.~\eqref{eq:eBS} depends only on the (spatial) Cartesian coordinates of the particles, $H$ is the positive-energy projected two-electron Hamiltonian with instantaneous (Coulomb or Coulomb--Breit) interaction ($I$), 
\begin{align}
  H=h_1+h_2 + \mL_{++} I \mL_{++} \; ,
  \label{eq:npham}
\end{align}
$h_i=c\bos{\alpha}_i\bos{p}_i+\beta_i m_i c^2 + U 1^{[4]}$ $(i=1,2)$ is the one-particle Dirac Hamiltonian in which $U$ can account for an external static Coulomb field (if there is any), and $\Lambda_{++}$ projects to the positive-energy (electronic) subspace of the $h_1+h_2$ non-interacting two-fermion problem.
For short, we call $H$ the no-pair Dirac--Coulomb (DC) or Dirac--Coulomb--Breit (DCB) Hamiltonian.

Pair corrections, retardation, and radiative corrections are included in the $H_\Delta$ term, Eq.~\eqref{eq:eBS} \cite{sucherPhD1958,DoKr74,Zhang96}. Contribution of $H_\Delta$ to atomic and molecular energies (QED) can be expected to be small, and hence, it can be treated as perturbation to the no-pair Hamiltonian. 

This framework offers a perturbative approach based on a \emph{relativistic} reference, alternative to earlier work using a non-relativistic reference state. 
Evaluation of the already formulated perturbative correction with $H_\Delta$ is left for future developments, which appears to be possible along the lines reviewed in Ref.~\cite{MaFeJeMa22}.
Although analytic evaluation of the energy and its corrections is not possible in this framework, the numerical results can be converged to high precision, which is demonstrated in the present work.

To compute no-pair, two-particle bound states, let us start with defining overall, center-of-mass, $R^\mu=(T,\bR)$, and relative, $r^\mu=(t,\br)$, covariant space-time coordinates as
\begin{align}
  R^\mu
  =
  \frac{m_1}{m_1+m_2} r_1^{\mu}+\frac{m_2}{m_1+m_2}r_2^{\mu} 
  \label{eq:comcoor}
\end{align}
and 
\begin{align}
  r^\mu 
  &= 
  r_1^{\mu}-r_2^{\mu} \; .
  \label{eq:relcoor}
\end{align}
Then, following Salpeter and Bethe \cite{SaBe51},
the wave function of an isolated system can be factorized as
\begin{align}
  \phi(r_1,r_2)=\eem^{-\iim P_\nu R^\nu} \Phi(r^\mu) \ 
  \label{eq:ansatz}  
\end{align}
with the total four-momentum, $P^\nu=(E,\bos{P})$.
By choosing the zero-total-momentum frame, $\bos{P}=\bos{0}$, we obtain
\begin{align}
  \phi(r_1,r_2)=\eem^{-\iim E T} \Phi(r^\mu) \;,
  \label{eq:zeroMomAnsatz}
\end{align}
where $E$ is the total energy of the system.
It is important to note that $\Phi(r^\mu)$, which describes the internal motion, depends on $r^\mu=(t,\br)$\emph{, i.e.,} not only on the $\br$ relative coordinates, but also on the $t$ relative time of the particles.
Fourier transformation with respect to this relative time variable yields the relative-energy dependent wave function
\begin{align}
  \tilde\Phi(\epsi,\br) 
  = 
  \int_{-\infty}^\infty \frac{\dd t}{(2\pi)^{1/2}} 
  \eem^{-\iim \epsi t} \Phi(t,\br)  \; .
\end{align}
In the exact equal-time formalism of Salpeter \cite{Sa52} and Sucher \cite{sucherPhD1958}, 
the equal-time ($t=0$) wave function appears, which depends only on the spatial coordinates, 
\begin{align}
\Psi(\br) = \int_{-\infty}^\infty \dd  \epsi \,  \tilde\Phi(\epsi,\br) \; ,
\end{align}
and the relative-energy dependence of the problem is accounted for in $H_\Delta$ in Eq.~\eqref{eq:eBS} \cite{sucherPhD1958}.

%
To obtain the Hamiltonian for the relative motion,
the chain rule for the coordinate transformation, Eqs.~(\ref{eq:comcoor}) and (\ref{eq:relcoor}), is used, and
it is also considered that  contribution from terms containing $\grad_{\bos{R}}$ 
vanishes due to the Eq.~(\ref{eq:ansatz}) choice of the ansatz for an isolated system and our choice of a $\bos{P}=\bos{0}$ zero-momentum-frame description, Eq.~(\ref{eq:zeroMomAnsatz}). 
Hence, the spatial momentum operators in this framework can be replaced according to
\begin{align}
  \bp_1 = -\iim\grad_1 \rightarrow \bp= -\iim \grad 
  \quad\quad\text{and}\quad\quad
  \bp_2 = -\iim\grad_2 \rightarrow -\bp = \iim \grad \, ,
  \label{eq:relativemom}
\end{align}
where $\grad(=\grad_{\bos{r}})$ collects the partial derivatives with respect to the $\bos{r}$ relative displacement vector components.
This simple replacement `rule' can be used to construct expressions for the relative motion from the two-particle expressions \cite{JeFeMa21,JeFeMa22,FeJeMa22,FeJeMa22b}.
As a result, the no-pair Dirac--Coulomb--Breit Hamiltonian for the relative motion is obtained as
\begin{align}
  H(1,2) = 
  { \footnotesize
  \mL_{\pupu}
  \left(%
    \begin{array}{@{} c@{\ \ }c@{\ \ }c@{\ \ }c @{}}
       V\unitfour & 
       -c \bsigma\four_2 \cdot \bp & 
       c \bsigma\four_1 \cdot \bp & 
       B\four \\
       -c\bsigma\four_2 \cdot \bp  & 
       V\unitfour-2m_2c^2\unitfour & 
       B\four & 
       c \bsigma\four_1 \cdot \bp \\
       c\bsigma\four_1 \cdot\bp & 
       B\four &
       V\unitfour-2m_1c^2\unitfour & 
       -c \bsigma\four_2 \cdot \bp \\
       B\four & 
       c \bsigma\four_1 \cdot \bp &
       -c \bsigma\four_2 \cdot \bp & 
       V\unitfour-2m_{12}c^2\unitfour \\
    \end{array}
  \right)
  \mL_{\pupu}
  }
  \label{eq:fullHam}
\end{align}
with $m_{12}=m_1+m_2$, $\bp = -\iim(\pd{}{r_{x}},\pd{}{r_{y}},\pd{}{r_{z}})$,
$\bsigma\four_1=(\sigma_x\otimes\unittwo,\sigma_y\otimes\unittwo,\sigma_z\otimes\unittwo)$
and
$\bsigma\four_2=(\unittwo\otimes \sigma_x,\unittwo\otimes\sigma_y,\unittwo\otimes\sigma_z)$, where $\sigma_x,\sigma_y,$ and $\sigma_z$ are the $2\times 2$ Pauli matrices.
We note that the operator in Eq.~(\ref{eq:fullHam}) contains a $-2m_ic^2$ shift ($i=1,2$) to match the non-relativistic energy scale.
Furthermore, the Coulomb interaction,  
\begin{align}
  V=\frac{q_1q_2}{r} 
\end{align}
is along the diagonal, whereas the Breit interaction, 
\begin{align}
  B\four
  =
  -q_1q_2
  \left[%
    \frac{1}{r}\bsigma_1\four\cdot\bsigma_2\four 
    -\frac{1}{2}
    \left\lbrace
      \left(%
        \bsigma_1\four\cdot\grad
      \right)\left(%
        \bsigma_2\four\cdot\grad
      \right) r  
    \right\rbrace
  \right] \; 
  \label{eq:breit}
\end{align}
can be found on the anti diagonal of the Hamiltonian.

The $\Lambda_{\pupu}$ positive-energy projector in Eq.~(\ref{eq:fullHam}) corresponds to the positive-energy (`electronic') states of 
the `bare', non-interacting Hamiltonian, \emph{i.e.,} 
Eq.~(\ref{eq:fullHam}) without $\Lambda_{\pupu}$ and without the $V1^{[4]}$ and $B^{[4]}$ interaction blocks.
Although the $\Lambda_{\pupu}$ free-particle projector in momentum space has an analytic form \cite{HaSu84}, we constructed it numerically in coordinate space by computing the eigenstates of the bare, non-interacting Hamiltonian over the space spanned by the basis functions used for the interacting computation. The positive-energy states were identified with the simple energy cutting approach (which can be checked by the complex scaling procedure) \cite{JeFeMa22}.

The no-pair Dirac--Coulomb and Dirac--Coulomb--Breit Hamiltonians are bounded from below (the positive-energy block, which is considered in this work, is decoupled from the rest), hence the $H\Psi=E\Psi$ wave equation can be solved using the variational procedure.

For a single particle, the (four-component) wave function is conveniently partitioned to large (l, first two) and small (s, last two) components. A good basis representation must fulfill a simple symmetry relation, which is necessary to provide a correct matrix representation (Mx) for the $\text{Mx}(p)\text{Mx}(p)=\text{Mx}(p^2)$ identity~\cite{ScWa82}.
The simplest implementation of this relation is provided by the (restricted) kinetic balance (KB) condition \cite{Ku84,Liu10}, 
\begin{align}
    \varphi^\text{s}=\frac{\bsigma^{[2]}\cdot \bp }{2mc} \varphi^\text{l} 
    \label{eq:kb}
\end{align}
for the basis function of the $\varphi^\text{s}$ small and $\varphi^\text{l}$ large components. 
Two(many)-particle relativistic quantities can be constructed 
with the block-wise (also called Tracy--Singh) direct product \cite{TrSi72,LiShLi12,ShLiLi17,SiMaRe15,JeFeMa21,JeFeMa22,FeJeMa22,FeJeMa22b}, which allows us to retain the large-small block structure, used already to write Eq.~(\ref{eq:fullHam}). 
The corresponding two-particle function, with highlighting the large (l) and small (s) component blocks, is
\begin{align}
    \bos{\varphi}
    =\left( \begin{array}{c}
         \varphi^\text{ll} \\
         \varphi^\text{ls} \\
         \varphi^\text{sl} \\
         \varphi^\text{ss}
    \end{array} \right)    \; .
\end{align}
For a variational procedure, we used the simplest two-particle generalization of the one-particle kinetic balance, Eq.~(\ref{eq:kb}), and implemented it in the sense of a transformation or metric \cite{Ku84,JeFeMa21,JeFeMa22,FeJeMa22,FeJeMa22b}:
\begin{align}
    H_\text{KB}=X^\dagger H X \; , 
    && 
    X=\text{diag}\left(%
      1\four, -\frac{\left(\bsigma_2\four\cdot \bp \right)}{2m_2c},%
      \frac{\left(\bsigma_1\four\cdot \bp \right)}{2m_1c},%
      -\frac{\left(\bsigma_1\four\cdot \bp \right)\left(\bsigma_2\four\cdot \bp \right)}{4m_1m_2c^2}\right) \; .
    \label{eq:XHX} 
\end{align}
We also note that the $X$ balance matrix used in this work can be `obtained' from the balance used for the Born--Oppenheimer systems \cite{JeFeMa21,JeFeMa22,FeJeMa22,FeJeMa22b} through the $\bp_1 \rightarrow \bp$ and $\bp_2 \rightarrow -\bp$ replacement, Eq.~(\ref{eq:relativemom}). The fundamental `guiding principle' for our construction of the two-particle balance has been solely to have a correct matrix representation of the $\text{Mx}(p)\text{Mx}(p)=\text{Mx}(p^2)$ identity, since the positive-energy projected Hamiltonian is bounded from below.
The transformed DCB Hamiltonian is 
\begin{align}
    H_\text{KB}=X^\dagger H(1,2) X=
    \begin{pmatrix}
    D_{1}\four & \frac{\bp^2}{2m_2}\unitfour & \frac{\bp^2}{2m_1}\unitfour & B_1\four \\
    \frac{\bp^2}{2m_2}\unitfour & D_{2}\four & B_2\four & \frac{\bp^4}{8c^2m_1m_2^2}\unitfour \\
    \frac{\bp^2}{2m_1}\unitfour & B_3\four & D_{3}\four & \frac{\bp^4}{8c^2m_1^2m_2}\unitfour \\
    B_4\four & \frac{\bp^4}{8c^2m_1m_2^2}\unitfour & \frac{\bp^4}{8c^2m_1^2m_2}\unitfour & D_{4}\four
    \end{pmatrix}
\end{align}
with the diagonal blocks, 
\begin{align}
    D_{1}\four&=V\unitfour\\
    \label{eq:DCHdiag1}
    D_{2}\four&=\frac{(\bsigma_2\cdot\bp)V\unitfour(\bsigma_2\cdot\bp)}{4m_2^2c^2}-\frac{\bp^2}{2m_2}\unitfour \\
    D_{3}\four&=\frac{(\bsigma_1\cdot\bp)V\unitfour(\bsigma_1\cdot\bp)}{4m_1^2c^2}-\frac{\bp^2}{2m_1}\unitfour \\
    D_{4}\four&=\frac{(\bsigma_1\cdot\bp)(\bsigma_2\cdot\bp)V\unitfour(\bsigma_1\cdot\bp)(\bsigma_2\cdot\bp)}{16m_1^2m_2^2c^4}-\frac{m_{12}\bp^4}{8m_1^2m_2^2c^2} \unitfour\; ,
    \label{eq:DCHdiag}
\end{align}
and the anti-diagonal blocks including the Breit interaction, Eq.~(\ref{eq:breit}),  
\begin{align}
    B_1\four&=-\frac{B\four(\bsigma_1\cdot\bp)(\bsigma_2\cdot\bp)}{4c^2m_1m_2} \\
    B_2\four&=-\frac{(\bsigma_2\cdot\bp_2)B\four(\bsigma_1\cdot\bp)}{4c^2m_1m_2}\\
    B_3\four&=-\frac{(\bsigma_1\cdot\bp)B\four(\bsigma_2\cdot\bp)}{4c^2m_1m_2}\\
    B_4\four&=-\frac{(\bsigma_2\cdot\bp)(\bsigma_1\cdot\bp)B\four}{4c^2m_1m_2}  \; .
\end{align}
The identity in the $X$-KB metric is
\begin{align}
    I_\text{KB}
    =X^\dagger X =\text{diag}\left(
    \unitfour,
    \frac{\bp^2}{4c^2m_2^2}\unitfour,
    \frac{\bp^2}{4c^2m_1^2}\unitfour,
    \frac{\bp^4}{16c^4m_1^2m_2^2}\unitfour
    \right) \; .
\end{align}
Then, the sixteen-component wave function is written as a linear-combination of spinor functions,
\begin{align}
  \Psi(\br) 
  &=
  \sum_{i=1}^{\nb} \sum_{\chi=1}^{16} c_{i\chi} f_i(\br) \bos{d}_{\chi} \; ,
\end{align}
where the $\bos{d}_\chi$ spinor basis vectors are sixteen-dimensional unit vectors, $(\bos{d}_\chi)_\rho=\delta_{\chi\rho}$ ($\chi,\rho=1,\ldots,16$). 
%
For the $f_i$ spatial functions, we use spherically symmetric Gaussian functions ($S^\texte$, $L=0$ orbital angular momentum and $p=+1$ even (e) parity),
\begin{align}
  f_i(\br)
  &= \eem^{-\zeta_i r^2} \; 
     \label{eq:gaussbasis}
\end{align}
with $\zeta_i > 0$ (to ensure square integrability).
We optimized the $\zeta_i$ Gaussian exponents ($i=1,\ldots,\nb$) by minimization of the non-relativistic $^1S^\texte$ ground-state energy to a p$\Eh{}(=10^{-12}\ \Eh{})$ precision range using quadruple precision arithmetic. 
Convergence of the non-relativistic and relativistic energies with respect to the basis size is shown in Table~\ref{tab:convergence}.
For selected systems and basis sizes, we continued the optimization of the $\zeta_i$ parameters by minimization of the no-pair DC(B) energy, and the computation remained variationally stable, the energy `converged from above'. (This variationally stable behaviour was absent during minimization of the relevant energy level of the bare DC Hamiltonian.)
We also note that there are no triplet contributions to the ground state ($1\  ^1S^\texte_0$) (p.~419 of Ref.~\cite{landau4}), since even-parity $^3P^\texte$ states do not exist for a pseudo-one-particle system (in contrast to helium-like systems \cite{JeMa22}).

In addition to variational no-pair DC and DCB computations, we computed the first-order perturbative Breit correction to the $n$th DC energy (with $n=1$ in this work)  by   \cite{FeJeMa22,FeJeMa22b}
\begin{align}
  E_{\text{DC}\langle \text{B}\rangle,n}
  =
  E_{\text{DC},n}
  +
  \langle 
    \Psi_{\text{DC},n} | X^\dagger B X \Psi_{\text{DC},n}
  \rangle 
  \label{eq:dcpB}
\end{align}
where $B$ is a sixteen-dimensional matrix with the $B^{[4]}$ blocks on its anti diagonal. The second-order perturbative Breit correction is computed as
\begin{align}    
    E_{\text{DC}\mathcal{B}_{2,n}}
    &=
    E_{\text{DC}\langle \text{B}\rangle,n}
    +
    \sum_{i \neq n} 
    \frac{%
      \abs{%
        \langle%
          \Psi_{\text{DC},i} 
          | X^\dagger B X 
          \Psi_{\text{DC},n}
        \rangle
      }^2
    }{%
      E_{\text{DC},i}-E_{\text{DC},n}
    }  
    \label{eq:pt2}
    \; .
\end{align}

The outlined algorithm has been implemented in the QUANTEN computer program, which is used as a molecular physics `platform' for pre-Born--Oppenheimer, non-adiabatic, upper- and lower-bound, perturbative- and variational relativistic developments \cite{Ma19review,FeMa19EF,FeMa19HH,FeKoMa20,FeMa22H3,IrJeMaMaRoPo21,RoJeMaPo22,MaFe22,FeMa22bethe,JeIrFeMa22,JeFeMa21,JeFeMa22,FeJeMa22,FeJeMa22b,FeMa22bethe,JeMa22}.
Throughout this work Hartree atomic units are used, and the speed of light is $c=\alpha^{-1}a_0\Eh/\hbar$ with $\alpha^{-1}=137.$035~999~084  \cite{codata18}.

All computed no-pair energies are listed in Table~\ref{tab:convergence}, their change with the basis size can be used to assess their convergence. Further minimization tests for no-pair the DC(B) energy did not reveal major changes.


%
For direct comparison of the computed no-pair energies with the current state-of-the-art nrQED values, we have (numerically) determined the $\alpha$ dependence of the no-pair energies. For this reason, we repeated the no-pair computations using the $\left\lbrace \alpha^{-1} \in \alpha_0^{-1} \pm n\ |\ n \in \left\lbrace -50,\ldots , 51 \right\rbrace\right\rbrace$ series of the interaction constant, where $\alpha_0$ labels the value taken from Ref.~\cite{codata18}. 
Then, we fitted the function 
\begin{align}
  F(\alpha) 
  = 
  \epsi_0 + \alpha^2 \varepsilon_2 + 
  \alpha^3 \varepsilon_3 + 
  \alpha^4 \ln(\alpha) \varepsilon_4' + 
  \alpha^4 \varepsilon_4  \; 
  \label{eq:fit}
\end{align}
to the series of the no-pair energies. Inclusion of higher-order, \emph{e.g.,} $\alpha^5$ and $\alpha^5 \ln \alpha$, terms in Eq.~\eqref{eq:fit} did not make any visible difference at the current numerical precision.
A small fitting error was obtained, which had orders of magnitude smaller root-mean-squared deviation than the estimated energy convergence, Table~\ref{tab:convergence}, and a smooth convergence of the fitted coefficients was observed with respect to the basis set size (Tables~S2--S5). 
To obtain consistent results, it was essential to include also the $\alpha^4 \ln \alpha$ term in Eq.~\eqref{eq:fit}, a simple $\alpha$ polynomial was insufficient to represent the high-precision no-pair energies (Table~\ref{tab:convergence}). This feature reveals a non-regular $\alpha$ depdendence of the no-pair energy \cite{HaSu84}, which is different from the known regular behaviour of an unprojected DC(B) equation \cite{FuMa54} (that is known to be inconsistent with Feynman's propagator \cite{Fe49a,Fe49b}).

Table~\ref{tab:compare} shows the comparison of the $\alpha$-dependence of the no-pair energies (fitted coefficients) and the nrQED corrections that were readily available to us or we could obtain with short calculation (\som). Excellent agreement is observed. The numerical deviation of the perturbative and fitted variational values is on the order of the convergence error of the no-pair energies (Table~\ref{tab:convergence}). The list of all coefficients fitted according to Eq.~\eqref{eq:fit} is provided in Table~S6. Tables~S2--S5 can be used to assess the convergence of these values with respect to the basis size.

Regarding the large mass, $m_2\rightarrow\infty$, limit  and comparison with the one-electron Dirac energy, it is necessary to 
consider that the (bare) one-electron Dirac equation is \emph{with}-pair (and correct for one electron). 
At $\alpha^3\Eh$ order, the one-electron Dirac limit is recovered from our no-pair computations, by appending the no-pair energy with the (one) pair correction.  
For $m_2\rightarrow \infty$, the one-pair Coulomb correction, Eq.~(3.9) of Ref.~\citenum{FuMa54}, is
\begin{align}
  E^{(3)}_{\text{C}_1}(m_1,\infty)
  &=
  \lim_{m_2\rightarrow \infty}
  E^{(3)}_{\text{C}_1}(m_1,m_2)
  \nonumber \\
  &=
  \lim_{m_2\rightarrow \infty}
  \frac{2\mu^3}{3\pi}
  \left(
    \frac{2}{m_1^2} - \frac{1}{m_1m_2} + \frac{2}{m_2^2}
  \right) 
  =
  \frac{4m_1}{3\pi}   
  \; .
  \label{eq:onepair}
\end{align}
In Table~\ref{tab:largemass}, we can (numerically) observe that the large $m_2$ limit of the $\varepsilon_3$ coefficient, obtained from fitting $F(\alpha)$ to the no-pair energies, converges to $-E^{(3)}_{\text{C}_1}(1,\infty)$, and hence, cancels with the pair corrections (the two-pair contribution, Eq.~(S12),  vanishes) for $m_2\rightarrow\infty$. 
Thereby, the one-particle Dirac limit is recovered at order $\alpha^3\Eh$. These properties emerge as simple consequence of using a two-particle relativistic wave equation obtained from the full relativistic QED theory.
It is also worth noting that the Breit contribution vanishes as $m_2\rightarrow \infty$ (Table~S7).

\vspace{1cm}
In this work, a computational relativistic quantum electrodynamics approach was put forward based on the exact equal-time Bethe--Salpeter equation.
It is demonstrated that a relativistic reference state can be converged to a sub-parts-per-billion relative precision by variational solution of the no-pair Dirac--Coulomb(--Breit) wave equation including the dominant, instantaneous part of the electromagnetic interaction.
The $\alpha$ fine-structure dependence of the computed energies are in excellent agreement with the formal non-relativistic QED results corresponding to polynomial and logarithmic corrections in $\alpha$, up to $\alpha^6\ln\alpha$ order in natural units ($\alpha^4\ln\alpha\Eh$) and reveal a non-regular nature of the $\alpha$ expansion about the non-relativistic reference.
Perturbative retardation, radiative, and pair corrections to the no-pair relativistic states had been formulated long ago~\cite{sucherPhD1958,DoKr74,Zhang96}, and their evaluation with the high-precision relativistic reference states computed in this work will be carried out in subsequent work.


\vspace{0.5cm}
\begin{acknowledgments}
%
Financial support of the European Research Council through a Starting Grant (No.~851421) is gratefully acknowledged. DF thanks a doctoral scholarship from the ÚNKP-22-4 New National Excellence Program of the Ministry for Innovation and Technology from the source of the National Research, Development, and Innovation Fund (ÚNKP-22-4-I-ELTE-51).
\end{acknowledgments}

\clearpage

\begin{table}[h]
    \centering
    \caption{%
      Convergence of the no-pair Dirac--Coulomb(--Breit) energies, in $\Eh$, computed in this work. The spatial basis, Eq.~\eqref{eq:gaussbasis}, used in the relativistic computation was parameterized by (numerical) minimization of the non-relativistic energy, $E_\text{nr}$.
      The numerical value for the analytic ($\infty$) non-relativistic energy is shown for reference.
      \label{tab:convergence}
    }
\scalebox{0.85}{%
    \begin{tabular}{@{} c@{\ } d{4.14}@{\ } d{4.14}@{\ } d{4.14}@{\ }  d{4.14}@{\ } d{4.14} @{}}
        \hline\hline \\[-0.25cm]
        $\nb$  & \multicolumn{1}{c}{$E_\text{nr}$} 
        & \multicolumn{1}{c}{$E_\text{DC}$} 
        & \multicolumn{1}{c}{$E_{\text{DC}\langle\text{B}\rangle}$}
        & \multicolumn{1}{c}{$E_{\text{DC}\mathcal{B}_2}$} 
        & \multicolumn{1}{c}{$E_\text{DCB}$}
        \\ \hline\\[-0.25cm]
        \multicolumn{6}{l}{Ps\ ($m_2/m_1=1)$: }  \\
	    10	&	-0.249\ 999\ 665\ 988\ 4	&	-0.249\ 997\ 227\ 989   &   -0.250\ 016\ 969\ 603	& -0.250\ 016\ 992\ 755	& -0.250\ 016\ 992\ 809 	\\
	    20	&	-0.249\ 999\ 999\ 919\ 4	&	-0.249\ 997\ 552\ 650	&	-0.250\ 017\ 362\ 124   & -0.250\ 017\ 403\ 806 & -0.250\ 017\ 404\ 023  	\\
	    30	&	-0.249\ 999\ 999\ 996\ 8	&	-0.249\ 997\ 552\ 766	&	-0.250\ 017\ 362\ 426   & -0.250\ 017\ 404\ 153 & -0.250\ 017\ 404\ 371  	\\
	    40	&	-0.249\ 999\ 999\ 999\ 6	&	-0.249\ 997\ 552\ 778	&	-0.250\ 017\ 362\ 470   & -0.250\ 017\ 404\ 205 & -0.250\ 017\ 404\ 425  	\\
	    50	&	-0.249\ 999\ 999\ 999\ 9	&	-0.249\ 997\ 552\ 780	&	-0.250\ 017\ 362\ 477   & -0.250\ 017\ 404\ 214 & -0.250\ 017\ 404\ 433  	\\
	    $\infty$	&	-0.250\ 000\ 000\ 000\ 0	&						&				            & 	& \\
        \hline\\[-0.25cm]        
        \multicolumn{6}{l}{Mu\ ($m_2/m_1=206.7682830)$:}  \\
        10	&	-0.497\ 592\ 269\ 419\ 4	&	-0.497\	598\ 739\ 220	&   -0.497\ 599\ 489\ 904  & -0.497\ 599\ 489\ 917 & -0.497\ 599\ 489\ 918	\\
        20	&	-0.497\ 593\ 472\ 285\ 4	&	-0.497\	600\ 024\ 240	&   -0.497\ 600\ 780\ 916  & -0.497\ 600\ 780\ 959 & -0.497\ 600\ 780\ 959	\\
        30	&	-0.497\ 593\ 472\ 874\ 8	&	-0.497\	600\ 025\ 977	&   -0.497\ 600\ 782\ 839  & -0.497\ 600\ 782\ 891 & -0.497\ 600\ 782\ 891	\\
        40	&	-0.497\ 593\ 472\ 910\ 8	&	-0.497\	600\ 026\ 241	&   -0.497\ 600\ 783\ 176  & -0.497\ 600\ 783\ 235 & -0.497\ 600\ 783\ 235	\\
        50	&	-0.497\ 593\ 472\ 915\ 7	&	-0.497\	600\ 026\ 282	&   -0.497\ 600\ 783\ 233  & -0.497\ 600\ 783\ 295 & -0.497\ 600\ 783\ 295	\\
        $\infty$	&	-0.497\ 593\ 472\ 917\ 1	&						&                          &	& \\
        \hline\\[-0.25cm]
        \multicolumn{6}{l}{H\ ($m_2/m_1=1836.15267343)$:}  \\
        10	&	-0.499\ 727\ 019\ 644\ 9 	&	-0.499\ 733\ 723\ 658   &   -0.499\ 733\ 809\ 460  & -0.499\ 733\ 809\ 460 & -0.499\ 733\ 809\ 460	\\
        20	&	-0.499\ 727\ 839\ 067\ 5 	&	-0.499\ 734\ 617\ 695   &   -0.499\ 734\ 704\ 007  & -0.499\ 734\ 704\ 008 & -0.499\ 734\ 704\ 008	\\
        30	&	-0.499\ 727\ 839\ 669\ 3 	&	-0.499\ 734\ 619\ 508   &   -0.499\ 734\ 705\ 842  & -0.499\ 734\ 705\ 843 & -0.499\ 734\ 705\ 843	\\
        40	&	-0.499\ 727\ 839\ 706\ 0 	&	-0.499\ 734\ 619\ 795   &   -0.499\ 734\ 706\ 138  & -0.499\ 734\ 706\ 139 & -0.499\ 734\ 706\ 140	\\
        50	&	-0.499\ 727\ 839\ 710\ 9 	&	-0.499\ 734\ 619\ 840   &   -0.499\ 734\ 706\ 186  & -0.499\ 734\ 706\ 187 & -0.499\ 734\ 706\ 187	\\
        $\infty$	&	-0.499\ 727\ 839\ 712\ 4	&						&  &				        	& \\
        \hline\\[-0.25cm]
        \multicolumn{6}{l}{\muH\ ($m_2/m_1=8.88024337):$}  \\
        10	 &	-92.920\ 263\ 579\ 73     &    -92.920\ 730\ 693\ 26  &  -92.923\ 396\ 814\ 39 & -92.923\ 397\ 816\ 36 & -92.923\ 397\ 817\ 07 \\
        20	 &	-92.920\ 416\ 825\ 53     &    -92.920\ 890\ 799\ 40  &  -92.923\ 572\ 907\ 75 & -92.923\ 575\ 558\ 50 & -92.923\ 575\ 566\ 96 \\
        30 	 &	-92.920\ 417\ 297\ 88     &    -92.920\ 891\ 278\ 83  &  -92.923\ 573\ 403\ 13 & -92.923\ 576\ 058\ 44 & -92.923\ 576\ 066\ 97 \\
        40	 &	-92.920\ 417\ 310\ 07     &    -92.920\ 891\ 312\ 69  &  -92.923\ 573\ 493\ 64 & -92.923\ 576\ 164\ 19 & -92.923\ 576\ 173\ 06 \\
        50	 &	-92.920\ 417\ 311\ 03     &    -92.920\ 891\ 313\ 65  &  -92.923\ 573\ 494\ 58 & -92.923\ 576\ 165\ 15 & -92.923\ 576\ 174\ 01 \\
        $\infty$   &  -92.920\ 417\ 311\ 31 & & & & \\        
        \hline\hline
    \end{tabular}
}
\end{table}

\begin{table}[h]
    \centering
    \caption{%
      Comparison of variational no-pair results and nrQED corrections. The 
      $F(\alpha)=\epsi_0 + \alpha^2 \varepsilon_2 + \alpha^3 \varepsilon_3 + \alpha^4 \ln(\alpha) \varepsilon_4' + \alpha^4 \varepsilon_4$ function was fitted to the no-pair energies to obtain the coefficients (var-fit). 
      All values correspond to Hartree atomic units. (All coefficients are listed in Table~S6.)
      \label{tab:compare}
    }
    \scalebox{0.88}{%
    \begin{tabular}{@{}l@{} %
      @{}c@{\ } d{3.7}d{3.7}d{3.7} %
      @{}c@{\ } d{3.7}d{3.7} %
      @{}c@{\ } d{3.7}d{3.6} @{}}
        \hline\hline \\[-0.35cm]
        & ~ &
        \multicolumn{3}{c}{DC} & ~ & 
        \multicolumn{2}{c}{DC$\langle\text{B}\rangle$} & ~ &
        \multicolumn{2}{c}{DCB} \\
        \cline{3-5}\cline{7-8}\cline{10-11} \\[-0.4cm]
        && 
        \multicolumn{1}{c}{$\varepsilon_2$} & 
        \multicolumn{1}{c}{$\varepsilon_3$} &
        \multicolumn{1}{c}{$\varepsilon_4'$} 
        && 
        \multicolumn{1}{c}{$\varepsilon_2$} & 
        \multicolumn{1}{c}{$\varepsilon_3$} 
        &&
        \multicolumn{1}{c}{$\varepsilon_2$} &
        \multicolumn{1}{c}{$\varepsilon_3$}        
        \\ \hline \\[-0.35cm]
        %
        \multicolumn{10}{l}{Ps $=\pos$:} \\
        var-fit 
        && 
           0.046\ 875 & -0.128\ 8  & -0.063\ 4 
        &&
          -0.328\ 125 & 0.280\ 2 
        &&
          -0.328\ 125 & 0.189\ 9 \\
        nrQED $^\text{a}$ 
        && 
           0.046\ 875 & -0.128\ 8  & -0.062\ 5
        && 
          -0.328\ 125 &  0.280\ 3 
        &&
          -0.328\ 125 &  \\
        $\alpha^n(\delta\varepsilon_n)\ ^\text{b}$ 
        && 
         -4.5 \cdot 10^{-12} & 7.2 \cdot 10^{-12} & 2.6 \cdot 10^{-12}
        && 
         -2.3\cdot 10^{-11} & 5.5\cdot 10^{-11} 
        && 
          2.3 \cdot 10^{-11} &  \\
        \hline\\[-0.35cm]
        \multicolumn{10}{l}{Mu $=\muo$:} \\
        var-fit 
        &&
         -0.120\ 227 & -0.419\ 3  & -0.967\ 2 
        && 
         -0.134\ 526 & -0.407\ 1 
        &&
         -0.134\ 526 & -0.407\ 2 \\
        nrQED $^\text{a}$ 
        &&
         -0.120\ 227 & -0.419\ 3  &
        && 
         -0.134\ 528 &  
        &&
         -0.134\ 528 &  \\
        $\alpha^n(\delta\varepsilon_n)\ ^\text{b}$ 
        &&
         -4.7 \cdot 10^{-11} & -1.2\cdot 10^{-11}  &
        && 
         -1.0\cdot 10^{-10} & 
        && 
         -1.1 \cdot 10^{-10} & \\
        \hline\\[-0.35cm]
        \multicolumn{10}{l}{H $=\hyd$:} \\
        var-fit 
        &&
         -0.124\ 455 & -0.423\ 8  & -0.983\ 7
        && 
         -0.126\ 086  & -0.422\ 4 
        &&
         -0.126\ 086  & -0.422\ 4 \\
        nrQED $^\text{a}$
        && 
         -0.124\ 456 & -0.423\ 8  &
        &&
         -0.126\ 087 & 
        &&
         -0.126\ 087 &  \\
        $\alpha^n(\delta\varepsilon_n)\ ^\text{b}$ 
        &&
         -6.1 \cdot 10^{-11} & -1.0\cdot 10^{-11}  &
        &&
         -6.8\cdot10^{-11} & 
        && 
         -6.8 \cdot 10^{-11} & \\
        \hline\\[-0.35cm]
        \multicolumn{10}{l}{$\mu$H $=\muhyd$:} \\
        var-fit 
        &&
         -8.437\ 67 & -67.886 & -130.550\ 2
        && 
         -59.154\ 212 & -18.860\ 6  
        &&
         -59.154\ 120 & -24.865\ 6 \\
        nrQED $^\text{a}$
        && 
         -8.437\ 70  &  -67.899   &
        &&
        -59.154\ 516 &   
        &&
        -59.154\ 516 &  \\
        $\alpha^n(\delta\varepsilon_n)\ ^\text{b}$ 
        && 
        -1.7\cdot 10^{-9} & -5.4\cdot 10^{-9}  &
        &&
        -1.6\cdot 10^{-8} & 
        &&  
        -2.1\cdot 10^{-8} & \\                  
        \hline \hline
    \end{tabular}
    }
    \begin{flushleft}
    {\footnotesize%
      $^\text{a}$~%
        The nrQED expressions and the corresponding literature references \cite{sucherPhD1958,FuMa54,landau4,KhMiYe93,Zhang96} are collected in the \som. \\
      $^\text{b}$~%
        $\alpha^n(\delta\varepsilon_n)$, in $\Eh$, with
        the $\delta\varepsilon_n=E^{(n)}-\varepsilon_n$ difference of the nrQED value and the fitted coefficient.
    }
    \end{flushleft}
\end{table}

\begin{table}
\caption{%
Large mass, $m_2\rightarrow \infty$, limit, of the $\alpha^3\Eh$-order fitted coefficient of the no-pair DC energy, Eq.~\eqref{eq:fit}. ($m_1=1$ corresponds to the electron mass.)
\label{tab:largemass}
}
\begin{tabular}{@{}ll d{5.9} d{3.8}@{}}
\hline\hline\\[-0.35cm]
  && 
  \multicolumn{1}{c}{$m_2$} & 
  \multicolumn{1}{c}{$\varepsilon_3$} \\ 
  \hline \\[-0.35cm]
  Ps  & $=\pos$ & 1              & -0.128\ 8 \\
  Mu  & $=\muo$ & 206.7682830    & -0.419\ 3 \\
  H   & $=\hyd$ & 1836.15267343  & -0.423\ 8 \\
  10H & $=\lbrace \text{e}^-, 10\text{p}^+ \rbrace$ & 18361.5267343  & -0.424\ 3 \\
  \hline\\[-0.35cm]
  \multicolumn{2}{c}{$-E^{(3)}_{\text{C}_1}(1,m_2)$ Eq.~\eqref{eq:onepair}} & 
  \multicolumn{1}{c}{$\infty$} & 
  -0.424\ 413... \\
\hline\hline \\  
\end{tabular}
\end{table}

\clearpage

\setcounter{section}{0}
\renewcommand{\thesection}{S\arabic{section}}
\setcounter{subsection}{0}
\renewcommand{\thesubsection}{S\arabic{section}.\arabic{subsection}}

\setcounter{equation}{0}
\renewcommand{\theequation}{S\arabic{equation}}

\setcounter{table}{0}
\renewcommand{\thetable}{S\arabic{table}}

\setcounter{figure}{0}
\renewcommand{\thefigure}{S\arabic{figure}}

~\\[0.cm]
\begin{center}
\begin{minipage}{0.8\linewidth}
\centering
\textbf{Supplementary Material} \\[0.25cm]

\textbf{%
Pre-Born--Oppenheimer Dirac--Coulomb--Breit computations for two-body systems
}
\end{minipage}
~\\[0.5cm]
\begin{minipage}{0.6\linewidth}
\centering

D\'avid Ferenc$^1$ and Edit M\'atyus$^{1,\ast}$ \\[0.15cm]

$^1$~\emph{ELTE, Eötvös Loránd University, Institute of Chemistry, 
Pázmány Péter sétány 1/A, Budapest, H-1117, Hungary} \\[0.15cm]
$^\ast$ edit.matyus@ttk.elte.hu \\
\end{minipage}
~\\[0.15cm]
(Dated: January 31, 2022)
\end{center}

~\\[1cm]
\begin{center}
\begin{minipage}{0.9\linewidth}
\noindent %
Contents: \\
S1. Non-relativistic QED expressions compiled from the literature \\
S2. Matrix elements \\
S3. Expectation values and mass-dependent correction formulae \\
S4. Convergence tables \\
S5. Fitted coefficients \\
~~~References
\end{minipage}
\end{center}

\clearpage

\section{Non-relativistic QED expressions compiled from the literature \label{sec:pt}}
In the non-relativistic QED approach, the energy is obtained by evaluating $E^{(n)}$ terms for increasing powers of $\alpha$ as corrections to the $E^{(0)}_\text{nr}$ non-relativistic energy (which is of $\alpha^0$ order in hartree atomic units),
\begin{align}
    E = E_\text{nr}^{(0)}+\alpha^2E^{(2)} + \alpha^3 E^{(3)} +\ldots \; 
    \label{eq:pt}
\end{align}
The Schrödinger equation of two-particle systems has a closed analytic solution, and the ground-state energy reads as
\begin{align}
  E_\text{nr}^{(0)}
  =
  -\frac{\mu}{2}  \; 
  \label{eq:nonrel}
\end{align}
with the reduced mass 
\begin{align}
  \mu=\frac{m_1m_2}{m_1+m_2} \; .
  \label{eq:redmass}
\end{align}
The next, (non-vanishing) $\alpha^2\Eh$-order correction is the sum of two terms arising from the Coulomb and the Breit (non-retarded part of transverse) interactions,
\begin{align}
    \alpha^2 E_\text{DCB}^{(2)}
    = \alpha^2  E_\text{DC}^{(2)} 
    + \alpha^2  E_\text{B}^{(2)} \;  ,
    \label{eq:E2CB}
\end{align}
which is obtained by calculating the expectation value of the following operators \cite{landau4},
\begin{align}
  H_\text{DC}^{(2)} 
  &=
  -\frac{1}{8}\left(\frac{1}{m_1^3}+\frac{1}{m_2^3}\right)(\bp^2)^2 -\frac{\pi}{2}\left(\frac{1}{m_1^2}+\frac{1}{m_2^2}\right)\delta(\br) 
  \label{eq:H2C}
  \\
  H_\text{B}^{(2)} 
  &=   
  -\frac{1}{2m_1m_2r}\left[\bp^2+\frac{\br(\br\bp)\bp}{r^2}\right] 
    -\frac{2\pi}{m_1m_2}\delta(\br) 
    \label{eq:H2B}
\end{align}
with the non-relativistic ground-state wave function.
These expectation values can be written in a closed, analytic form (Secs.~\ref{sec:mx} and \ref{sec:exptval}).
For the $1\ ^1S^\text{e}$ state, they are 
\begin{align}
    E_\text{DC}^{(2)}
    &=
    E_\text{MV} + E_\text{D}  \; 
\end{align}
with
\begin{align}
    E_\text{MV} = -\frac{5}{8} \mu ^4 \left(\frac{1}{m_1^3}+\frac{1}{m_2^3}\right) &&
    E_\text{D} = \frac{\mu^3}{2}  \left(\frac{1}{m_1^2}+\frac{1}{m_2^2}\right)   \; ,
\end{align}
from the first and second terms of Eq.~\eqref{eq:H2C}, respectively, and
\begin{align}
    E_\text{B}^{(2)} 
    &=    E_\text{oo} +  E_\text{ss} \; ,
\end{align}
where
\begin{align}
    E_\text{oo} = -\frac{ \mu ^3}{m_1m_2} \quad\quad\text{and}\quad\quad
    E_\text{ss} =  -\frac{ 2\mu ^3}{m_1m_2}  \; 
\end{align}
from the two terms of Eq.~\eqref{eq:H2B}. 

\begin{figure}
    \centering
    \includegraphics[width=0.8\textwidth]{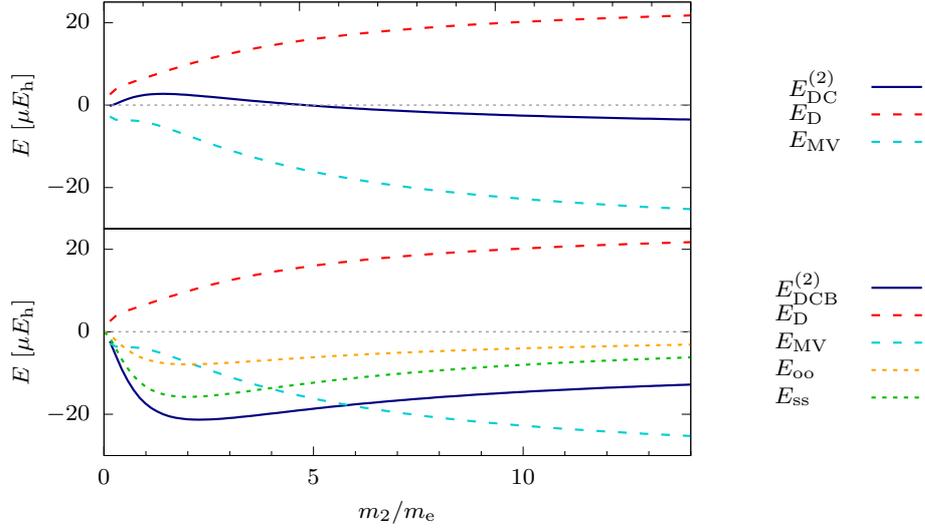}
    \caption{
    Dependence of the $\alpha^2\Eh $-order Dirac--Coulomb (top) and Dirac--Coulomb--Breit (bottom) energy corrections of two-particle systems on the $m_2$ particle mass with $m_1=1 (m_\text{e})$.  
    For $m_1=1$, the $E^{(2)}_\text{DC}$ correction vanishes for the 
    $m_2=0.208\ 71$ and 4.791\ 29 values,
    \emph{i.e.,} up to $\alpha^2\Eh$ the DC relativistic energy equals the non-relativistic energy.
    }
    \label{fig:E2m}
\end{figure}

Figure~\ref{fig:E2m} shows the $m_2$ dependence of the corrections for the $m_1=1$ case.
Interestingly, the relativistic DC correction vanishes for the $m_2=0.209$ and $m_2=4.791$ values.
In contrast to the  non-relativistic energy, the mass-dependence of the corrections is not only through the reduced mass of the constituent particles.

The correction arising from non-crossed photons at $\alpha^3\Eh$ order has been reported by Fulton and Martin (Eq.~(3.7) of Ref.~\cite{FuMa54}),
\begin{align}
   E^{(3)}_{\text{C}_{0,2}}(m_1,m_2)
   &= 
   -\frac{2\mu^3}{3\pi}\left( \frac{2}{m_1^2}+\frac{1}{m_1m_2}+\frac{2}{m_2^2}\right)  \; .
    \label{eq:E3C02p}
\end{align}
We note that this expression contains the sum of the no-pair and the two-pair corrections (indicated by the `0,2' subscript). To the best of our knowledge, the single and two-pair Coulomb corrections separately do not have any simple form for general $m_1,m_2$ masses, but can be calculated from the integral (two-pair part of Eqs.~(3.1a)--(3.6) of Ref.~\cite{FuMa54}):
\begin{align}
    E_{\text{C}_2}^{(3)}(m_1,m_2)
    =
    -\frac{2\mu^3}{\pi} \int_0^\infty 
    \dd k\;
    \frac{(E_1(k)-m_1)(E_2(k)-m_2)}{E_1(k)+E_2(k)+m_1+m_2} \quad\text{with}\quad E_i(k) = \sqrt{m_i^2+k^2} \; .
    \label{eq:E3C2p}
\end{align}
This integral can be evaluated by using (for example) a symbolic algebra program, and the resulting (lengthy) expression can be evaluated for the selected $m_1$ and $m_2$ masses.
For the special case of $m_1=m_2=1$, the integral simplifies to, Eq.~(4.26b) of Ref.~\cite{sucherPhD1958},
\begin{align}
  E_{\text{C}_2}^{(3)}(1,1)
  =
  -\frac{1}{8\pi}
  \left(
  \frac{5}{3}
  -\frac{\pi}{2}
  \right)\; ,
\end{align}
which together with using the simple expression for the non-crossed photon correction of Fulton and Martin, Eq.~(\ref{eq:E3C02p}), can be used to obtain the third-order perturbative no-pair Coulomb correction for unit masses, $m_1=m_2=1$:
\begin{align}
  E_{\text{C}_0}^{(3)}(1,1)
  =
  E_{\text{C}_{0,2}}^{(3)}(1,1)
  -
  E_{\text{C}_2}^{(3)}(1,1)
  =
  -\frac{1}{8\pi}
  \left(
  \frac{\pi}{2}
  +\frac{5}{3}
  \right)  
  \approx
  -0.128\ 815
  \; ,
  \label{eq:E3C0p}
\end{align}
which is the same as the third-order no-pair Coulomb correction reported by Sucher (for the two electrons in helium, Eq.~(3.99) of Ref.~\cite{sucherPhD1958}). 
The zero- plus two-pair contribution, Eq.~(\ref{eq:E3C02p}), for unit masses is 
\begin{align}
  E^{(3)}_{\text{C}_{0,2}}(1,1)
  &= 
  -\frac{5}{12\pi} 
  \approx
  -0.132\ 629 \; .
\end{align}
The $E^{(3)}_{\text{C}_0}$ no-pair contribution was not separately reported in the literature for non-unit masses, and 
we calculated it for the relevant mass values using Eqs.~\eqref{eq:E3C02p} and \eqref{eq:E3C2p} (Table~\ref{tab:corr}).

The $\alpha^3\Eh{}$-order contribution from a single instantaneous Breit photon exchange including also the Coulomb ladder is known for unit masses (positronium) \cite{sucherPhD1958}, 
\begin{align}
    E^{(3)}_{\text{B}}(1,1)=\frac{1}{2\pi}\left(1+\frac{\pi}{2}\right) \approx 0.409\ 155 \; .
\end{align}

As to the logarithmic contributions, the nrQED expansion of the no-pair DC energy does not contain any $\alpha^3\ln \alpha$-order term, but there are $\alpha^4\ln \alpha$-order contributions. 
The analytic nrQED value of this $\alpha^4\ln \alpha$-order correction can be easily calculated for the Coulomb exchange and positronium using 
Eq. (39) of Ref.~\cite{KhMiYe93}, which gives $-\frac{1}{16}=-0.062\ 5$ for the ground state. Our value fitted in the largest basis set is $\varepsilon'_4=-0.063\ 4$. 
Further logarithmic contributions for unequal masses and for transverse photon exchange are discussed in Refs.~\cite{KhMiYe93} and \cite{Zhang96}. 

In nrQED, the $\alpha^3\ln\alpha$-order logarithmic contributions are attributed to the `usual' infrared divergence of QED (and are obtained as non-relativistic radiative corrections), whereas the $\alpha^4\ln\alpha$-order contribution is of `relativistic nature' \cite{KhMiYe93}, related to the `relativistic momentum' range \cite{Zhang96}.

From our point of view, the various (\emph{e.g.,} logarithmic) corrections can be understood as consequence of the \emph{mathematical structure} created by the $\alpha$ expansion about the non-relativistic reference ($\alpha^0$ order).

\begin{table}
\caption{%
    Non-relativistic energy and perturbative correction values, in Hartree atomic units, calculated using the analytic expressions, Eqs.~\eqref{eq:pt}--\eqref{eq:E3C0p}, compiled from the literature \cite{landau4,FuMa54,sucherPhD1958}. 
    The $m_2/m_1$ mass ratios are taken from Ref.~\cite{codata18}.
  \label{tab:corr}
}
\scalebox{0.79}{%
\begin{tabular}{@{}l@{}c d{4.8} d{4.15} d{3.8}  d{3.8}  d{3.8} d{3.8} d{3.8} @{}}
 \hline\hline\\[-0.35cm]    
  &   & 
  \multicolumn{1}{c}{$m_2/m_1$} & 
  \multicolumn{1}{c}{$E^{(0)}_\text{nr}$} & 
  \multicolumn{1}{c}{$E^{(2)}_\text{DC}$} &
  \multicolumn{1}{c}{$E^{(2)}_\text{DCB}$} &
  \multicolumn{1}{c}{$E^{(3)}_{\text{C}_{0,2}}$}  &
  \multicolumn{1}{c}{$E^{(3)}_{\text{C}_{2}}$}  &
  \multicolumn{1}{c}{$E^{(3)}_{\text{C}_0}$}  \\
  \hline\\[-0.35cm]
  Ps & $=\pos$ & 1
     & -0.250\ 000\ 000\ 000   &  0.046\ 875   & -0.328\ 125  & -0.132\ 629  & -0.003\ 815 & -0.128\ 815^\text{a} \\
  Mu & $=\muo$ & 206.7682830
     & -0.497\ 593\ 472\ 917   &  -0.120\ 227  & -0.134\ 528  & -0.419\ 336  & -3.6\cdot 10^{-6} & -0.419\ 332 \\
  H  & $=\hyd$ & 1836.15267343
     & -0.499\ 727\ 839\ 712   &  -0.124\ 456  & -0.126\ 087  & -0.423\ 836  & -4.7\cdot 10^{-8} & -0.423\ 836\\
  \muH{} & $=\muhyd$ & 8.88024337
     &  -92.920\ 417\ 311\ 307 & -8.437\ 699   & -59.154\ 516 & -68.110\ 857 & -0.210\ 9268 & -67.899\ 930 \\     
  \hline\hline  
\end{tabular}
}
\begin{flushleft}
{\footnotesize
  $^\text{a}$~%
    Eq.~(\ref{eq:E3C0p}) was used for Ps \cite{sucherPhD1958}. 
}    
\end{flushleft}
\end{table}

\section{Matrix elements \label{sec:mx}}
\noindent The spatial basis functions are
\begin{align}
  f_\mu(\bos{r})
  =
  \left(\frac{2\zeta_\mu}{\pi}\right)^{3/4}e^{-\zeta_\mu r^2} = N_\mu e^{-\zeta_\mu r^2} \; .
\end{align}
The following notation is used, $i,j,k,l\in \{x,y,z\}$ denote Cartesian components
\begin{align}
    \zeta_{\mu\nu}&=\zeta_\mu+\zeta_\nu \\
    \int \dd^3\br&=\int_{-\infty}^\infty \dd x\, \dd y \, \dd z \\
    \int \dd^3\br\ e^{-\zeta r^2}&=\left(\frac{\pi}{\zeta}\right)^{3/2}\\
    r&=\sqrt{x^2+y^2+z^2} \\
    N_\mu&=\left(\frac{2\zeta_\mu}{\pi}\right)^{3/4} \; .
\end{align}
In what follows, the Einstein summation convention is understood for the $i,j,k,l$ Cartesian indices. 
The derivatives of the basis functions are
\begin{align}
  \partial_i e^{-\zeta_\mu r^2}&=-2\zeta_\mu r_ie^{-\zeta_\mu r^2}\label{eq:d1phi}\\
    \partial_j\partial_ie^{-\zeta_\mu r^2}&=\left[-2\zeta_\mu\delta_{ij}+4\zeta_\mu^2r_ir_j\right]e^{-\zeta_\mu r^2} \\
    \partial_k\partial_j\partial_ie^{-\zeta_\mu r^2}&=\left[ 4\zeta_\mu^2(\delta_{ij}r_k+\delta_{ik}r_j+\delta_{kj}r_i)-8\zeta_\mu^3r_ir_jr_k\right]e^{-\zeta_\mu r^2} \; .
    \label{eq:d3phi}
\end{align}
A useful integrals for the calculation of the Coulomb matrix elements include
\begin{align}
    \int_0^\infty \dd t\, (a+t^2)^{-3/2}  &= \frac{1}{a} \\
    \int_0^\infty \dd t\, (a+t^2)^{-5/2}  &= \frac{1}{3a^2} \\
    \int_0^\infty \dd t\, (a+t^2)^{-7/2}  &= \frac{1}{15a^3} \; .
\end{align}

\begin{align}
    \braket{f_\mu}{f_\nu}
    =N_\mu N_\nu\int \dd^3\br\  e^{-(\zeta_\mu+\zeta_\nu) r^2}= \frac{(4\zeta_\mu\zeta_\nu)^{3/4}}{\zeta_{\mu\nu}^{3/2}} 
\end{align}

\begin{align}
    \matrixel{f_\mu}{V}{f_\nu}
    =&N_\mu N_\nu\int \dd^3\br\ \frac{1}{r}e^{-\zeta_{\mu\nu}r^2}=N_\mu N_\nu\frac{2}{\sqrt{\pi}}\int_0^\infty \dd t \int \dd^3\br\ \frac{1}{r}e^{-(\zeta_{\mu\nu}+t^2)r^2}  \nonumber \\
    &=N_\mu N_\nu\frac{2}{\sqrt{\pi}}\int_0^\infty \dd t\frac{\pi^{3/2}}{(\zeta_{\mu\nu}+t^2)^{3/2}} = N_\mu N_\nu \frac{2\pi}{\zeta_{\mu\nu}} \nonumber \\
    &=\sqrt{\frac{32}{\pi}}\frac{(\zeta_\mu\zeta_\nu)^{3/4}}{\zeta_{\mu\nu}}
\end{align}

\begin{align}
    \matrixel{f_\mu}{\grad^2}{f_\nu}
    =&N_\mu N_\nu\int \dd^3\br\ \left[-2\delta_{ii}\zeta_\mu+4\zeta_\mu^2r_ir_i\right]e^{-\zeta_{\mu\nu}r^2}  \nonumber \\
    &=\frac{-6\zeta_\mu(4\zeta_\mu\zeta_\nu)^{3/4}}{\zeta_{\mu\nu}^{3/2}}+12\frac{ \sqrt{2}\zeta_\mu^{11/4}\zeta_\nu^{3/4}}{\zeta_{\mu\nu}^{5/2}}
\end{align}

\begin{align}
    \matrixel{f_\mu}{\grad^2\grad^2}{f_\nu}
    =&N_\mu N_\nu  \int \dd^3\br\ \left[ -2\zeta_\mu \delta_{ii}+4\zeta_\mu^2r_ir_i\right]\left[ -2\zeta_\nu \delta_{jj}+4\zeta_\nu^2r_jr_j\right] e^{-\zeta_{\mu\nu}r^2} \nonumber\\
    &=N_\mu N_\nu \left[4\zeta_\mu\zeta_\nu\delta_{ii}\delta_{jj} \int \dd^3\br\ e^{-\zeta_{\mu\nu}r^2} \nonumber  \right. \\
    &-8\zeta_\mu\zeta_\nu^2\delta_{ii}\int \dd^3\br\ r_j r_j e^{-\zeta_{\mu\nu}r^2} \nonumber \\
    &-8\zeta_\nu\zeta_\mu^2\delta_{jj}\int \dd^3\br\ r_i r_i e^{-\zeta_{\mu\nu}r^2} \nonumber \\
    &+16\zeta_\nu^2\zeta_\mu^2 \left.\int \dd^3\br\ r_j r_j r_i r_i e^{-\zeta_{\mu\nu}r^2}\right]
\end{align}
\begin{align}
    36N_\mu N_\nu\zeta_\mu\zeta_\nu \int \dd^3\br\ e^{-\zeta_{\mu\nu}r^2}= 36N_\mu N_\nu \zeta_\mu\zeta_\nu\pi^{3/2}\zeta_{\mu\nu}^{-3/2}
\end{align}
\begin{align}
    -24N_\mu N_\nu\zeta_\mu\zeta_\nu^2 \int \dd^3\br\ r_j r_j e^{-\zeta_{\mu\nu}r^2}=-36N_\mu N_\nu\zeta_\mu\zeta_\nu^2\pi^{3/2}\zeta_{\mu\nu}^{-5/2}
\end{align}
The sum of the first three terms is zero:
\begin{align}
    &\frac{36N_\mu N_\nu \zeta_\mu\zeta_\nu\pi^{3/2}}{\zeta_{\mu\nu}^{3/2}}%
    -\frac{36N_\mu N_\nu\zeta_\mu\zeta_\nu^2\pi^{3/2}}{\zeta_{\mu\nu}^{5/2}}%
    -\frac{36N_\mu N_\nu\zeta_\nu\zeta_\mu^2\pi^{3/2}}{\zeta_{\mu\nu}^{5/2}}\nonumber\\
    &=\frac{36N_\mu N_\nu \zeta_\mu\zeta_\nu\pi^{3/2}\zeta_{\mu\nu}}{\zeta_{\mu\nu}^{3/2}\zeta_{\mu\nu}}-\frac{36N_\mu N_\nu\zeta_\mu\zeta_\nu^2\pi^{3/2}}{\zeta_{\mu\nu}^{5/2}}%
    -\frac{36N_\mu N_\nu\zeta_\nu\zeta_\mu^2\pi^{3/2}}{\zeta_{\mu\nu}^{5/2}}\nonumber\\
    &=\frac{36N_\mu N_\nu \zeta_\mu\zeta_\nu\pi^{3/2}(\zeta_\mu+\zeta_\nu)}{\zeta_{\mu\nu}^{5/2}}-\frac{36N_\mu N_\nu\zeta_\mu\zeta_\nu^2\pi^{3/2}}{\zeta_{\mu\nu}^{5/2}}%
    -\frac{36N_\mu N_\nu\zeta_\nu\zeta_\mu^2\pi^{3/2}}{\zeta_{\mu\nu}^{5/2}}\nonumber\\
    &=\frac{36N_\mu N_\nu \zeta_\mu^2\zeta_\nu\pi^{3/2}}{\zeta_{\mu\nu}^{5/2}}+\frac{36N_\mu N_\nu \zeta_\mu\zeta_\nu^2\pi^{3/2}}{\zeta_{\mu\nu}^{5/2}}-\frac{36N_\mu N_\nu\zeta_\mu\zeta_\nu^2\pi^{3/2}}{\zeta_{\mu\nu}^{5/2}}-\frac{36N_\mu N_\nu\zeta_\nu\zeta_\mu^2\pi^{3/2}}{\zeta_{\mu\nu}^{5/2}}=0
\end{align}
\begin{align}
    16N_\mu N_\nu\zeta_\nu^2\zeta_\mu^2 \int \dd^3\br\ r^2 r^2 e^{-\zeta_{\mu\nu}r^2}= 60N_\mu N_\nu\zeta_\nu^2\zeta_\mu^2 \pi^{3/2}  \zeta_{\mu\nu}^{-7/2}=120\frac{\sqrt{2}\zeta_\mu^{11/4}\zeta_\nu^{11/4}}{\zeta_{\mu\nu}^{7/2}}
\end{align}

\begin{align}
    \matrixel{\partial_i f_\mu}{V}{\partial_j f_\nu}=N_\mu N_\nu\frac{8\pi}{3}\zeta_\mu\zeta_\nu\zeta_{\mu\nu}^{-2}=\sqrt{\frac{2}{\pi}}\frac{16}{3}\frac{\zeta_\mu^{7/4}\zeta_\nu^{7/4}}{\zeta_{\mu\nu}^2}\delta_{ij}
\end{align}
\begin{align}
    \matrixel{\partial_i\partial_j f_\mu}{V}{\partial_k\partial_l f_\nu}
    =&
    N_\mu N_\nu \left[ \vphantom{\frac{1}{1}}\right.\frac{8\pi\zeta_\mu\zeta_\nu}{\zeta_{\mu\nu}}\delta_{ij}\delta_{kl} \nonumber\\
    &-\frac{16\pi\zeta_\mu\zeta_\nu^2}{3\zeta_{\mu\nu}^{2}}\delta_{ij}\delta_{kl} %
    -\frac{16\pi\zeta_\nu\zeta_\mu^2}{3\zeta_{\mu\nu}^{2}}\delta_{ij}\delta_{kl} \nonumber \\
    &+\left.\frac{64\pi\zeta^2_k\zeta_\nu^2}{15\zeta_{\mu\nu}^{3}}\left(\delta_{il} \delta_{jk}+\delta_{ik} \delta_{jl}+\delta_{ij} \delta_{kl}\right)\right]
    \label{eq:pipjVpkpl}
\end{align}
It is convenient to introduce the following integrals
\begin{align}
     F_1(i,j)=&\int\dd^3\br\ \frac{1}{r} r_ir_j f_\mu f_\nu  \\
     &=N_\mu N_\nu \frac{1}{3} \pi \delta_{ij} \zeta_{\mu\nu}^{-2} 
     =\delta_{ij} \frac{4}{3}\sqrt{\frac{2}{\pi}}  \zeta_\mu^{3/4}\zeta_\nu^{3/4} \zeta_{\mu\nu}^{-2}
      \label{eq:pbobf1}
      \\
     F_2(i,j,k,l)=&\int\dd^3\br\ \frac{1}{r} r_ir_jr_kr_l f_\mu f_\nu \\
     &=N_\mu N_\nu \frac{2}{15} \pi\zeta_{\mu\nu}^{-3} \left(\delta_{il} \delta_{jk}+\delta_{ik} \delta_{jl}+\delta_{ij} \delta_{kl}\right)  \nonumber \\
     &=\frac{8}{15}\sqrt{\frac{2}{\pi}} \zeta_\mu^{3/4}\zeta_\nu^{3/4} \zeta_{\mu\nu}^{-3} \left(\delta_{il} \delta_{jk}+\delta_{ik} \delta_{jl}+\delta_{ij} \delta_{kl}\right) \; .
     \label{eq:pbobf2}
\end{align}
The Breit terms are expressed with the following functions
\begin{align}
    I_1(\mu,\nu,i,j,k,l) =& \int \dd^3\br\, \left(\partial_i f_\mu \right) \frac{r_j}{r}  \left(\partial_k\partial_l f_\nu \right) \nonumber \\
    &= 4\zeta_\mu\zeta_\nu \delta_{kl} F_1(i,j)-8\zeta_\mu\zeta_\nu^2F_2(i,j,k,l) \\
    I_2(\mu,\nu,i,j,k,l) =& \int \dd^3\br\,  f_\mu \frac{r_i}{r}  \left(\partial_j\partial_k\partial_l f_\nu\right) \nonumber \\
    &= 4\zeta_\nu^2\left[ \delta_{jk} F_1(i,l)
    +\delta_{jl} F_1(i,k)
    +\delta_{kl} F_1(i,j)
    \right] -8\zeta_\nu^3F_2(i,j,k,l) \; .
\end{align}
The matrix elements of the Breit operator are
\begin{align}
    \mxelmn{}{B_1}{}=&\frac{1}{4c^2m_1m_2}\mxelmn{}{\left\lbrace(\bsigma_1\cdot\grad)(\bsigma_2\cdot\grad)r \right\rbrace (\bsigma_1\cdot\grad)(\bsigma_2\cdot\grad)}{} \nonumber \\
    &= \frac{\sigma_{1_i}\sigma_{2_j}\sigma_{1_k}\sigma_{2_l}}{4c^2m_1m_2} \mxelmn{}{\left\lbrace \partial_i\partial_j r \right\rbrace \partial_k \partial_l }{} \nonumber \\
    &= \frac{\sigma_{1_i}\sigma_{2_j}\sigma_{1_k}\sigma_{2_l}}{4c^2m_1m_2} 
    \left[ 
    -\mxelmn{\partial_i}{\frac{r_j}{r}}{\partial_k\partial_l}
    -
    \mxelmn{}{\frac{r_j}{r}}{\partial_i\partial_k\partial_l}
    \right] \nonumber \\
    &= \frac{\sigma_{1_i}\sigma_{2_j}\sigma_{1_k}\sigma_{2_l}}{4c^2m_1m_2} 
    \left[ 
    -I_1(\mu,\nu,i,j,k,l)-I_2(\mu,\nu,j,i,k,l)
    \right]  \\
    \mxelmn{}{B_2}{}=&\frac{1}{4c^2m_1m_2}\mxelmn{}{(\bsigma_2\cdot\grad)
    \left\lbrace(\bsigma_1\cdot\grad)(\bsigma_2\cdot\grad)r \right\rbrace
    (\bsigma_1\cdot\grad)}{} \nonumber \\
    &=\frac{\sigma_{2_i}\sigma_{1_j}\sigma_{2_k}\sigma_{1_l}}{4c^2m_1m_2} \mxelmn{}{\partial_i \lbrace \partial_j\partial_k r \rbrace \partial_l}{} \nonumber \\
    &=-\frac{\sigma_{2_i}\sigma_{1_j}\sigma_{2_k}\sigma_{1_l}}{4c^2m_1m_2} \mxelmn{\partial_i}{\left\lbrace\partial_j\frac{r_k}r\right\rbrace}{\partial_l} \nonumber \\
    &=\frac{\sigma_{2_i}\sigma_{1_j}\sigma_{2_k}\sigma_{1_l}}{4c^2m_1m_2} \left[ \mxelmn{\partial_j\partial_i}{\frac{r_k}{r}}{\partial_l}
    +\mxelmn{\partial_i}{\frac{r_k}{r}}{\partial_j\partial_l} 
    \right] \nonumber \\
    &=\frac{\sigma_{2_i}\sigma_{1_j}\sigma_{2_k}\sigma_{1_l}}{4c^2m_1m_2} \left[
    I_1(\nu,\mu,l,k,i,j)+I_1(\mu,\nu,i,k,j,l)
    \right] \\
    \mxelmn{}{B_3}{}=&\frac{1}{4c^2m_1m_2}\mxelmn{}{(\bsigma_1\cdot\grad)
    \left\lbrace(\bsigma_1\cdot\grad)(\bsigma_2\cdot\grad)r \right\rbrace
    (\bsigma_2\cdot\grad)}{} \nonumber \\
    &=\frac{\sigma_{1_i}\sigma_{1_j}\sigma_{2_k}\sigma_{2_l}}{4c^2m_1m_2} \mxelmn{}{\partial_i \lbrace \partial_j\partial_k r \rbrace \partial_l}{} \nonumber \\
    &=-\frac{\sigma_{1_i}\sigma_{1_j}\sigma_{2_k}\sigma_{2_l}}{4c^2m_1m_2}
    \mxelmn{\partial_i}{\left\lbrace\partial_j\frac{r_k}r\right\rbrace}{\partial_l} \nonumber \\
    &=\frac{\sigma_{1_i}\sigma_{1_j}\sigma_{2_k}\sigma_{2_l}}{4c^2m_1m_2}
    \left[ \mxelmn{\partial_j\partial_i}{\frac{r_k}{r}}{\partial_l}
    +\mxelmn{\partial_i}{\frac{r_k}{r}}{\partial_j\partial_l} 
    \right] \nonumber \\
     &=\frac{\sigma_{1_i}\sigma_{1_j}\sigma_{2_k}\sigma_{2_l}}{4c^2m_1m_2}
    \left[
    I_1(\nu,\mu,l,k,i,j)+I_1(\mu,\nu,i,k,j,l)
    \right] \\
    \mxelmn{}{B_4}{}=&\frac{1}{4c^2m_1m_2}\mxelmn{}{(\bsigma_2\cdot\grad) (\bsigma_1\cdot\grad)
    \left\lbrace(\bsigma_1\cdot\grad)(\bsigma_2\cdot\grad)r \right\rbrace
   }{} \nonumber \\
   &= \frac{\sigma_{2_i}\sigma_{1_j}\sigma_{1_k}\sigma_{2_l}}{4c^2m_1m_2} \mxelmn{}{\partial_i \partial_j 
   \left\lbrace \partial_k\partial_l r \right\rbrace }{} \nonumber \\
   &= \frac{\sigma_{2_i}\sigma_{1_j}\sigma_{1_k}\sigma_{2_l}}{4c^2m_1m_2} \mxelmn{\partial_i \partial_j }{
   \left\lbrace \partial_k\frac{r_l}{r} \right\rbrace }{} \nonumber  \\
    &= \frac{\sigma_{2_i}\sigma_{1_j}\sigma_{1_k}\sigma_{2_l}}{4c^2m_1m_2} \left[
    -\mxelmn{\partial_k\partial_i\partial_j}{\frac{r_l}{r}}{}
    -\mxelmn{\partial_i\partial_j}{\frac{r_l}{r}}{\partial_k} 
    \right] \nonumber \\
    &= \frac{\sigma_{2_i}\sigma_{1_j}\sigma_{1_k}\sigma_{2_l}}{4c^2m_1m_2} \left[
    -I_2(\nu,\mu,l,i,j,k)-I_1(\nu,\mu,k,l,i,j)
    \right]
\end{align}

\section{Expectation values and mass-dependent correction formulae \label{sec:exptval}}
The necessary expectation values with arbitrary reduced mass are
\begin{align}
    E_{n,l} 
    &=- \frac{\mu}{2n^2} \\
    \matrixel{n,l}{r^{-1}}{n,l} &=\frac{\mu}{n^2} \\
    \matrixel{n,l}{r^{-2}}{n,l} &=\frac{2\mu^2}{n^3(2l+1)} \\
    \matrixel{n,l}{\delta(\br)}{n,l}
    &=|\psi_{n,l}(0)|^2
    = \frac{\mu^3}{\pi n^3}\delta_{l0}
\end{align}
The following operators contribute to the $\alpha^2\Eh{}$ DC energy
\begin{align}
    H_\text{MV}&=-\frac{1}{8}\left(\frac{1}{m_1^3}+\frac{1}{m_2^3}\right)(\bp^2)^2
    \\
    H_\text{D}&=-\frac{\pi}{2}\left(\frac{1}{m_1^2}+\frac{1}{m_2^2}\right)\delta(\br)  \; ,
\end{align}
and the additional terms from the Breit interaction (for $^1S$ states) are
\begin{align}
    H_\text{oo}&=-\frac{1}{2m_1m_2r}\left[\bp^2+\frac{\br(\br\bp)\bp}{r^2}\right] 
    \\
    H_\text{ss}&=-\frac{2\pi}{m_1m_2}\delta(\br) \; .
\end{align}
We use (pp.~421--422 of Ref.~\cite{landau4})
\begin{align}
    \langle\bp^2\rangle&=2\mu\left\langle\left(E+\frac{1}{r}\right) \right\rangle \\
    \langle(\bp^2)^2\rangle&= 4\mu^2 \left\langle\left(E+\frac{1}{r}\right)^2 \right\rangle +16\mu^2 \pi |\psi(0)|^2 -16\mu^2\pi |\psi(0)|^2\delta_{l0} \\
    \left\langle
    \frac{\bp^2}{r}+
    \frac{\br(\br\bp)\bp}{r^3} 
    \right\rangle&= 
    4\mu\left\langle\frac{1}{r}\left(E+\frac{1}{r}\right) \right\rangle-4\pi|\psi(0)|^2-l(l+1)
    \langle r^{-3} \rangle
\end{align}
to obtain the required expectation values. For $l=0$, we get
\begin{align}
    -\frac{1}{8}\left(\frac{1}{m_1^3}+\frac{1}{m_2^3}\right) \langle(\bp^2)^2\rangle 
    &= \frac{\mu^4}{2}\left(\frac{1}{m_1^3}+\frac{1}{m_2^3}\right)\left( \frac{3}{4n^4}-\frac{2}{n^3}\right) \\
    \frac{\pi}{2}\left(\frac{1}{m_1^2}+\frac{1}{m_2^2}\right) |\psi(0)|^2 
    &= \frac{\mu^3}{2n^3} \left(\frac{1}{m_1^2}+\frac{1}{m_2^2}\right) \\
    -\frac{1}{2m_1m_2}\left\langle\frac{1}{r}\left[\bp^2+\frac{\br(\br\bp)\bp}{r^2}\right] \right\rangle  
    &= \frac{1}{2m_1m_2}\left( 
    \frac{2\mu^3}{n^4} 
    -\frac{8\mu^3}{n^3}
    +\frac{4\mu^3}{n^3} \right) \\
    -\frac{2\pi}{m_1m_2}
    \langle \delta(\br) \rangle
    &= -\frac{2\mu^3}{m_1m_2n^3} \; .
\end{align}

\clearpage
\section{Convergence tables}
\noindent%
Convergence of the fitted $\varepsilon_0,\varepsilon_2,\varepsilon_3,\varepsilon'_4$ and $\varepsilon_4$ coefficients of Eq.~(30) with respect to the $\nb$ number of basis functions.

\begin{table}[h]
    \centering
    \caption{%
      Ps \label{tab:convPs}
    }
    \begin{tabular}{ l@{\ \ } c@{\ \ } d{3.15} d{3.8} d{3.6} d{3.6} d{3.6} }
        \hline\hline
           & $\nb$ & 
        \multicolumn{1}{c}{$\varepsilon_0$}   &
        \multicolumn{1}{c}{$\varepsilon_2$}   &
        \multicolumn{1}{c}{$\varepsilon_3$}   &
        \multicolumn{1}{c}{$\varepsilon_4'$}   &
        \multicolumn{1}{c}{$\varepsilon_4$}    \\ \hline
        \multirow{5}{*}{DC} 
         & 10 & -0.249\ 999\ 665\ 675 & 0.045\ 946 & 0.048\ 6 & 3.734\ 6 & 8.537\ 5 \\
        & 20 & -0.249\ 999\ 999\ 911 & 0.046\ 879 & -0.130\ 3 & -0.135\ 4 & -0.140\ 8 \\
        & 30 & -0.250\ 000\ 000\ 004 & 0.046\ 877 & -0.129\ 4 & -0.088\ 9 & 0.008\ 8 \\
        & 40 & -0.250\ 000\ 000\ 001 & 0.046\ 875 & -0.128\ 9 & -0.066\ 7 & 0.074\ 7 \\
        & 50 & -0.250\ 000\ 000\ 000 & 0.046\ 875 & -0.128\ 8 & -0.063\ 4 & 0.083\ 8 \\
        \cline{1-7} 
        \multirow{5}{*}{DC$\langle$B$\rangle$} 
        & 10 & -0.249\ 999\ 666\ 654 & -0.325\ 371 & -0.115\ 3 & -9.451\ 5 & -22.409\ 5 \\
        & 20 & -0.249\ 999\ 999\ 906 & -0.328\ 113 & 0.278\ 7 & -0.218\ 4 & -0.401\ 2 \\
        & 30 & -0.249\ 999\ 999\ 975 & -0.328\ 125 & 0.280\ 8 & -0.173\ 5 & -0.302\ 2 \\
        & 40 & -0.249\ 999\ 999\ 993 & -0.328\ 125 & 0.280\ 3 & -0.196\ 8 & -0.375\ 3 \\
        & 50 & -0.249\ 999\ 999\ 999 & -0.328\ 125 & 0.280\ 2 & -0.202\ 6 & -0.392\ 3 \\
        \cline{1-7} 
        \multirow{5}{*}{DCB}
        & 10 & -0.249\ 999\ 666\ 272 & -0.325\ 514 & -0.040\ 2 & -2.918\ 2 & -6.200\ 3 \\
        & 20 & -0.249\ 999\ 999\ 841 & -0.328\ 118 & 0.189\ 5 & 0.246\ 4 & -0.590\ 5 \\
        & 30 & -0.249\ 999\ 999\ 977 & -0.328\ 124 & 0.189\ 7 & 0.230\ 8 & -0.654\ 0 \\
        & 40 & -0.249\ 999\ 999\ 994 & -0.328\ 125 & 0.189\ 8 & 0.230\ 3 & -0.657\ 4 \\
        & 50 & -0.249\ 999\ 999\ 996 & -0.328\ 125 & 0.189\ 9 & 0.232\ 9 & -0.649\ 6 \\
        \hline\hline
	\end{tabular}
\end{table}

\begin{table}[h]
    \centering
    \caption{%
      Mu \label{tab:convMu}
    }
    \begin{tabular}{ l@{\ \ } c@{\ \ } d{3.15} d{3.8} d{3.6} d{3.6} d{3.6} }
        \hline\hline
        & $\nb$ & 
        \multicolumn{1}{c}{$\varepsilon_0$}   &
        \multicolumn{1}{c}{$\varepsilon_2$}   &
        \multicolumn{1}{c}{$\varepsilon_3$}   &
        \multicolumn{1}{c}{$\varepsilon_4'$}   &
        \multicolumn{1}{c}{$\varepsilon_4$}    \\ \hline
        \multirow{5}{*}{DC} 
 & 10 & -0.497\ 592\ 269\ 583 & -0.120\ 332 & 0.144\ 8 & 20.427\ 9 & 58.866\ 4 \\
 & 20 & -0.497\ 593\ 472\ 091 & -0.120\ 167 & -0.429\ 1 & -1.264\ 1 & -1.406\ 6 \\
 & 30 & -0.497\ 593\ 472\ 817 & -0.120\ 215 & -0.421\ 2 & -1.023\ 3 & -0.754\ 9 \\
 & 40 & -0.497\ 593\ 472\ 893 & -0.120\ 225 & -0.419\ 6 & -0.975\ 0 & -0.624\ 3 \\
 & 50 & -0.497\ 593\ 472\ 904 & -0.120\ 227 & -0.419\ 3 & -0.967\ 2 & -0.603\ 3 \\
        \cline{1-7} 
        \multirow{5}{*}{DC$\langle$B$\rangle$} 
 & 10 & -0.497\ 592\ 269\ 591 & -0.134\ 452 & 0.141\ 0 & 19.985\ 8 & 57.659\ 7 \\
 & 20 & -0.497\ 593\ 472\ 089 & -0.134\ 461 & -0.416\ 9 & -1.255\ 4 & -1.446\ 6 \\
 & 30 & -0.497\ 593\ 472\ 817 & -0.134\ 513 & -0.409\ 0 & -1.015\ 1 & -0.796\ 3 \\
 & 40 & -0.497\ 593\ 472\ 892 & -0.134\ 524 & -0.407\ 3 & -0.966\ 6 & -0.665\ 2 \\
 & 50 & -0.497\ 593\ 472\ 904 & -0.134\ 526 & -0.407\ 1 & -0.958\ 9 & -0.644\ 5 \\
        \cline{1-7} 
        \multirow{5}{*}{DCB} 
 & 10 & -0.497\ 592\ 269\ 590 & -0.134\ 452 & 0.141\ 0 & 19.988\ 0 & 57.663\ 3 \\
 & 20 & -0.497\ 593\ 472\ 089 & -0.134\ 461 & -0.417\ 0 & -1.253\ 9 & -1.445\ 0 \\
 & 30 & -0.497\ 593\ 472\ 817 & -0.134\ 513 & -0.409\ 1 & -1.013\ 6 & -0.794\ 5 \\
 & 40 & -0.497\ 593\ 472\ 893 & -0.134\ 524 & -0.407\ 5 & -0.965\ 0 & -0.662\ 8 \\
 & 50 & -0.497\ 593\ 472\ 904 & -0.134\ 526 & -0.407\ 2 & -0.957\ 3 & -0.641\ 9 \\
        \hline\hline
	\end{tabular}
\end{table}

\begin{table}[h]
    \centering
    \caption{%
      H \label{tab:convH}
    }
    \begin{tabular}{ l@{\ \ } c@{\ \ } d{3.15} d{3.8} d{3.6} d{3.6} d{3.6} }
        \hline\hline
        & $\nb$ & 
        \multicolumn{1}{c}{$\varepsilon_0$}   &
        \multicolumn{1}{c}{$\varepsilon_2$}   &
        \multicolumn{1}{c}{$\varepsilon_3$}   &
        \multicolumn{1}{c}{$\varepsilon_4'$}   &
        \multicolumn{1}{c}{$\varepsilon_4$}    \\ \hline
        \multirow{5}{*}{DC} 
 & 10 & -0.499\ 727\ 021\ 003 & -0.124\ 246 & 0.049\ 6 & 19.391\ 8 & 58.146\ 5 \\
 & 20 & -0.499\ 727\ 838\ 870 & -0.124\ 393 & -0.433\ 8 & -1.285\ 0 & -1.441\ 8 \\
 & 30 & -0.499\ 727\ 839\ 611 & -0.124\ 443 & -0.425\ 7 & -1.040\ 9 & -0.781\ 0 \\
 & 40 & -0.499\ 727\ 839\ 687 & -0.124\ 453 & -0.424\ 1 & -0.991\ 7 & -0.647\ 9 \\
 & 50 & -0.499\ 727\ 839\ 699 & -0.124\ 455 & -0.423\ 8 & -0.983\ 7 & -0.626\ 5 \\
        \cline{1-7} 
        \multirow{5}{*}{DC$\langle$B$\rangle$} 
 & 10 & -0.499\ 727\ 021\ 056 & -0.125\ 860 & 0.049\ 3 & 19.339\ 4 & 57.998\ 3 \\
 & 20 & -0.499\ 727\ 838\ 869 & -0.126\ 024 & -0.432\ 4 & -1.283\ 9 & -1.446\ 3 \\
 & 30 & -0.499\ 727\ 839\ 611 & -0.126\ 074 & -0.424\ 3 & -1.039\ 8 & -0.785\ 6 \\
 & 40 & -0.499\ 727\ 839\ 687 & -0.126\ 084 & -0.422\ 7 & -0.990\ 5 & -0.652\ 4 \\
 & 50 & -0.499\ 727\ 839\ 699 & -0.126\ 086 & -0.422\ 4 & -0.982\ 6 & -0.630\ 9 \\
        \cline{1-7} 
        \multirow{5}{*}{DCB} 
 & 10 & -0.499\ 727\ 021\ 001 & -0.125\ 860 & 0.049\ 3 & 19.339\ 9 & 57.999\ 3 \\
 & 20 & -0.499\ 727\ 838\ 869 & -0.126\ 024 & -0.432\ 4 & -1.283\ 9 & -1.446\ 2 \\
 & 30 & -0.499\ 727\ 839\ 611 & -0.126\ 074 & -0.424\ 3 & -1.039\ 8 & -0.785\ 6 \\
 & 40 & -0.499\ 727\ 839\ 687 & -0.126\ 084 & -0.422\ 7 & -0.990\ 5 & -0.652\ 4 \\
 & 50 & -0.499\ 727\ 839\ 699 & -0.126\ 086 & -0.422\ 4 & -0.982\ 6 & -0.630\ 9 \\
        \hline\hline
	\end{tabular}
\end{table}

\begin{table}[h]
    \centering
    \caption{%
      $\mu$H \label{tab:convmuH}
    }
    \begin{tabular}{ l@{\ \ } c@{\ \ } d{3.15} d{3.8} d{3.6} d{5.6} d{5.6} }
        \hline\hline
        & $\nb$ & 
        \multicolumn{1}{c}{$\varepsilon_0$}   &
        \multicolumn{1}{c}{$\varepsilon_2$}   &
        \multicolumn{1}{c}{$\varepsilon_3$}   &
        \multicolumn{1}{c}{$\varepsilon_4'$}   &
        \multicolumn{1}{c}{$\varepsilon_4$}    \\ \hline
        \multirow{5}{*}{DC} 
 & 10 & -92.920\ 263\ 622\ 900 & -8.580\ 427 & 21.237\ 3 & 3\ 178.582\ 7 & 9\ 149.317\ 3 \\
 & 20 & -92.920\ 416\ 817\ 900 & -8.436\ 173 & -68.216\ 1 & -142.593\ 7 & -79.362\ 0 \\
 & 30 & -92.920\ 417\ 290\ 800 & -8.436\ 688 & -68.097\ 8 & -138.442\ 5 & -67.746\ 6 \\
 & 40 & -92.920\ 417\ 309\ 200 & -8.437\ 667 & -67.885\ 9 & -130.535\ 5 & -44.976\ 0 \\
 & 50 & -92.920\ 417\ 310\ 100 & -8.437\ 667 & -67.886\ 2 & -130.550\ 2 & -45.021\ 0 \\
        \cline{1-7} 
        \multirow{5}{*}{DC$\langle$B$\rangle$} 
 & 10 & -92.920\ 263\ 643\ 100 & -58.741\ 568 & 7.760\ 3 & 1\ 575.605\ 5 & 4\ 889.224\ 2 \\
 & 20 & -92.920\ 416\ 808\ 100 & -59.150\ 377 & -19.385\ 6 & -125.924\ 0 & -171.285\ 6 \\
 & 30 & -92.920\ 417\ 286\ 500 & -59.150\ 638 & -19.459\ 2 & -131.205\ 7 & -188.305\ 9 \\
 & 40 & -92.920\ 417\ 307\ 300 & -59.154\ 213 & -18.860\ 2 & -113.954\ 0 & -142.929\ 1 \\
 & 50 & -92.920\ 417\ 308\ 300 & -59.154\ 212 & -18.860\ 6 & -113.969\ 5 & -142.976\ 5 \\
        \cline{1-7} 
        \multirow{5}{*}{DCB} 
 & 10 & -92.920\ 263\ 637\ 500 & -58.744\ 360 & 9.491\ 5 & 1\ 756.008\ 4 & 5\ 236.465\ 9 \\
 & 20 & -92.920\ 416\ 792\ 300 & -59.150\ 698 & -25.262\ 7 & -103.891\ 6 & -194.837\ 7 \\
 & 30 & -92.920\ 417\ 273\ 300 & -59.151\ 372 & -25.181\ 3 & -101.424\ 0 & -187.901\ 1 \\
 & 40 & -92.920\ 417\ 306\ 600 & -59.154\ 115 & -24.867\ 0 & -94.725\ 5 & -172.162\ 0 \\
 & 50 & -92.920\ 417\ 307\ 500 & -59.154\ 119 & -24.865\ 6 & -94.658\ 1 & -171.955\ 8 \\
        \hline\hline
	\end{tabular}
\end{table}

\clearpage
\section{Fitted coefficients}

\begin{table}[h]
    \centering
    \caption{%
      Coefficients of the fitted 
      $F(\alpha) 
       = 
       \epsi_0 + \alpha^2 \varepsilon_2 + 
       \alpha^3 \varepsilon_3 + 
       \alpha^4\ln (\alpha) \varepsilon'_4 + 
       \alpha^4 \varepsilon_4 $
       polynomial to the no-pair energies evaluated for a series of $\alpha$ values
       using the largest basis sets generated in this work.
       All values correspond to hartree atomic units.
    }
    \begin{tabular}{@{} l@{\ \ } d{4.14}@{\ } d{4.6}@{\ } d{4.5}@{\ }  d{4.5}@{\ } d{4.3} @{}}
        \hline\hline \\[-0.35cm]
        & \multicolumn{1}{c}{$\epsi_0$} &
        \multicolumn{1}{c}{$\epsi_2$} &
        \multicolumn{1}{c}{$\epsi_3$} &
        \multicolumn{1}{c}{$\epsi_4'$} &
        \multicolumn{1}{c}{$\epsi_4$} \\ \hline\\[-0.35cm]
        \multicolumn{6}{l}{Ps\ ($m_2/m_1=1)$: }  \\
$E_\text{DC}$ %
 & -0.250\ 000\ 000\ 000 & 0.046\ 875 & -0.128\ 8 & -0.063\ 4 & 0.084 \\
$E_{\text{DC}\langle\text{B}\rangle}$ %
 & -0.249\ 999\ 999\ 999 & -0.328\ 125 & 0.280\ 2 & -0.202\ 6 & -0.392 \\
$E_{\text{DC}\mathcal{B}_2}$ %
 & -0.249\ 999\ 999\ 993 & -0.328\ 126 & 0.190\ 3 & 0.271\ 1 & -0.425 \\
$E_\text{DCB}$ %
 &-0.249\ 999\ 999\ 996 & -0.328\ 125 & 0.189\ 9 & 0.232\ 9 & -0.650 \\
        \hline\\[-0.35cm]
        \multicolumn{6}{l}{Mu\ ($m_2/m_1=206.7682830)$:}  \\
$E_\text{DC}$ %
 & -0.497\ 593\ 472\ 904 & -0.120\ 227 & -0.419\ 3 & -0.967\ 2 & -0.603 \\
$E_{\text{DC}\langle\text{B}\rangle}$ %
 & -0.497\ 593\ 472\ 904 & -0.134\ 526 & -0.407\ 1 & -0.958\ 9 & -0.644 \\
$E_{\text{DC}\mathcal{B}_2}$ %
 & -0.497\ 593\ 472\ 904 & -0.134\ 526 & -0.407\ 2 & -0.957\ 3 & -0.642 \\
$E_\text{DCB}$ %
 &-0.497\ 593\ 472\ 904 & -0.134\ 526 & -0.407\ 2 & -0.957\ 3 & -0.642 \\
        \hline\\[-0.35cm]
        \multicolumn{6}{l}{H\ ($m_2/m_1=1836.15267343)$:}  \\
$E_\text{DC}$ %
 & -0.499\ 727\ 839\ 699 & -0.124\ 455 & -0.423\ 8 & -0.983\ 7 & -0.626 \\
$E_{\text{DC}\langle\text{B}\rangle}$ %
 & -0.499\ 727\ 839\ 699 & -0.126\ 086 & -0.422\ 4 & -0.982\ 6 & -0.631 \\
$E_{\text{DC}\mathcal{B}_2}$ %
 & -0.499\ 727\ 839\ 699 & -0.126\ 086 & -0.422\ 4 & -0.982\ 6 & -0.631 \\
$E_\text{DCB}$ %
 &-0.499\ 727\ 839\ 699 & -0.126\ 086 & -0.422\ 4 & -0.982\ 6 & -0.631 \\
        \hline\\[-0.35cm]
        \multicolumn{6}{l}{\muH\ ($m_2/m_1=8.88024337)$:}  \\
$E_\text{DC}$ %
 & -92.920\ 417\ 310\ 141 & -8.437\ 667 & -67.886\ 2 & -130.550\ 2 & -45.021 \\
$E_{\text{DC}\langle\text{B}\rangle}$ %
 & -92.920\ 417\ 308\ 281 & -59.154\ 212 & -18.860\ 6 & -113.969\ 5 & -142.977 \\
$E_{\text{DC}\mathcal{B}_2}$ %
 & -92.920\ 417\ 307\ 444 & -59.154\ 145 & -24.853\ 3 & -93.505\ 9 & -164.387 \\
$E_\text{DCB}$ %
 &-92.920\ 417\ 307\ 523 & -59.154\ 119 & -24.865\ 6 & -94.658\ 1 & -171.956 \\
        \hline\hline
    \end{tabular}
\end{table}

\begin{table}
  \centering
  \caption{%
    Relative importance, in ppm ($10^{-6}$), of the Dirac--Coulomb and Breit contributions with respect to the mass ratio of the two fermions. 
    \label{tab:relimport}
  }
  \begin{tabular}{@{} l@{}l c r d{4.4} d{4.4} d{4.4} @{}}
    \hline\hline \\[-0.25cm]  
    &&
    \multicolumn{1}{c}{$\frac{m_2}{m_1}$} &	
    \multicolumn{1}{c}{$\frac{E_{\DC}-E_\text{nr}}{|E_\text{nr}|}$} &	
    \multicolumn{1}{c}{$\frac{E_{\DCpB}-E_{\DC}}{|E_{\DC}|}$} &
    \multicolumn{1}{c}{$\frac{E_{\DCptwoB}-E_{\DCpB}}{|E_{\DCpB}|}$} &
    \multicolumn{1}{c}{$\frac{E_{\DCB}-E_{\DCptwoB}}{|E_{\DCptwoB}|}$}	\\
    \hline\\[-0.4cm]
    Ps & $=\pos$	    &	1	            &	   9.788\ 9	    &	-79.238\ 8	    &	-0.166\ 9	&	-0.000\ 9	\\
    $\mu$H & $=\muhyd$  &	8.88024337	    &	$-$5.101\ 2	    &	-28.865\ 4	    &	-0.028\ 7	&	-0.000\ 1	\\
    Mu & $=\muo$	    &	206.768283	    &	$-$13.170\ 1	    &	 -1.521\ 2	    &	-0.000\ 1	&	 0.000\ 0	\\
    H & $=\hyd$	        &	1836.15267343	&	$-$13.567\ 6$^\text{a}$	    &	 -0.172\ 8	    &  	 0.000\ 0	&	 0.000\ 0	\\
    \hline
    $\mu$H$_\infty$ & $=\lbrace \mu^-,\text{p}_\infty^+\rbrace$	     &	$\infty$	&	 $-$13.313\ 2    &	 0	    &  	 0	&	 0	\\    
    H$_\infty$ & $=\lbrace \text{e}^-,\text{p}_\infty^+\rbrace$	     &	$\infty$	&	 $-$13.313\ 2    &	 0	    &  	 0	&	 0	\\
    \hline\hline    
  \end{tabular}
  \begin{flushleft}
{\footnotesize%
    $^\text{a}$~%
    By adding the $\alpha^3\Eh$ one-pair Coulomb correction to the no-pair DC energy, Eq.~31, we obtain
    $(E_\text{DC}+E^{(3)}_{\text{C}_1}-E_\text{nr})/E_\text{nr}=-13.237\ 6$.
}
  \end{flushleft}
\end{table}

\clearpage
%

\end{document}